\documentclass[usenatbib]{mnras}
\usepackage{enumitem}

\usepackage{tabularx}
\usepackage{multirow}
\usepackage{pdflscape} %
\usepackage{graphicx}
\usepackage[space]{grffile}
\usepackage{latexsym}
\usepackage{amsfonts,amsmath,amssymb}
\usepackage{url}
\usepackage[utf8]{inputenc} 
\usepackage{hyperref}
\hypersetup{colorlinks=false,pdfborder={0 0 0}}
\usepackage{textcomp}
\usepackage{longtable}

\newcommand{\NLARGEMATCH}{753}
\newcommand{\NKNOWNLT}{3093}
\newcommand{\NMATCHED}{732}
\newcommand{\NMULTIMATCH}{ 21}
\newcommand{\NBRIGHT}{34}
\newcommand{\NLARGEDIST}{five}
\newcommand{\NINCAT}{695}
\newcommand{\NRP}{660}
\newcommand{\NWITHPI}{151}
\newcommand{\NSPTOPT}{543}
\newcommand{\NSPTNIR}{384}
\newcommand{\NSPTPHO}{69}
\newcommand{\NUNRESOLVED}{eight}

\newcommand{\NGOODWISEWA}{648}

\newcommand{\NGOODWISEWB}{649}

\newcommand{\NGOODWISEWC}{446}

\newcommand{\NBINPUBFOUND}{27}
\newcommand{\NBINNEWFOUND}{20}
\newcommand{\NBINSYSTOT}{47}
\newcommand{\NBINTOTFOUND}{100}

\newcommand{\G}{\textit{Gaia}}
\newcommand{\SMA}{Paper~1}
\newcommand{\Gcat}{GUCDScat}
\newcommand{\DR}{\textit{Gaia}~DR2}
\newcommand{\POSITIONOFPROX}{61}
\newcommand{\NUMINFIVEPC}{792}
\newcommand{\NUMWITHSIMBADINFIVEPC}{38}
\newcommand{\NUMWITHSIMBADOUTFIVEPC}{34}
\providecommand{\dt}[1]{{\tt #1}} 
\newcommand{\masyr}{\,mas\,yr$^{-1}$}
\newcommand{\gbp}{{G_\mathrm{BP}}}
\newcommand{\grp}{{G_\mathrm{RP}}}

\usepackage{natbib}

\title[GUCDS II: The end of the main sequence]{The  \G~Ultra-Cool
  Dwarf Sample -- II: Structure at the end of the main sequence} 

\author[Smart et al.]
{R. L. Smart$^{1}$\thanks{\tt E-mail: richard.smart@inaf.it}, 
F. Marocco $^{2}$,
L.~M.~Sarro$^{3}$,
D.~Barrado$^{4}$,
J. C. Beam\'in$^{5,6}$,\newauthor
J.~A.~Caballero$^{4}$,
H. R. A. Jones$^7$
\\
$^{1}$Istituto Nazionale di Astrofisica, Osservatorio Astrofisico di Torino, Strada Osservatorio 20, 10025 Pino Torinese, Italy\\
$^{2}$Jet Propulsion Laboratory, California Institute of Technology, 4800 Oak Grove Dr., Pasadena, CA 91109, USA\\
$^{3}$Departamento de Inteligencia Artificial, ETSI Inform\'atica, UNED, Juan del Rosal, 16 28040 Madrid, Spain\\
$^{4}$Centro de Astrobiolog\'ia (INTA-CSIC), ESAC campus, Camino Bajo del Castillo s/n, E-28692, Villanueva de la Ca\~nada, Madrid, Spain\\
$^{5}$Instituto de F\'isica y Astronom\'ia, Facultad de Ciencias, Universidad de Valpara\'iso, Ave. Gran Breta\~na 1111, Playa Ancha, Valpara\'iso, Chile.\\
$^{6}$N\'ucleo Astroqu\'imica y Astrofísica, Facultad de Ingenier\'ia, Universidad Aut\'onoma de Chile, Chile \\
$^{7}$School of Physics, Astronomy and Mathematics, University of Hertfordshire, College Lane, Hatfield AL10 9AB, UK
}

\begin{document}



\maketitle

\begin{abstract}

We identify and investigate known late M, L and T dwarfs in the \G\ second
data release. This sample is being used as a training set in the \G\ data
processing chain of the ultra-cool dwarfs work package.  We find
\NINCAT\ objects in the optical spectral range M8 to T6 with accurate
\G\ coordinates, proper motions, and parallaxes which we combine with
published spectral types and photometry from large area
optical and infrared sky surveys. We find that \NBINTOTFOUND\ objects are in
\NBINSYSTOT\ multiple systems, of which \NBINPUBFOUND\ systems are published
and \NBINNEWFOUND\ are new. These will be useful benchmark systems and
we discuss the requirements to produce a complete catalog of
multiple systems with an ultra-cool dwarf component.  We examine the
magnitudes in the \G\ passbands and find that the $\gbp$ magnitudes are
unreliable and should not be used for these objects. We examine progressively
redder colour-magnitude diagrams and see a notable increase in the main
sequence scatter and a bi-variate main sequence for old and young objects. We
provide an absolute magnitude -- spectral sub-type calibration for $G$ and
$\grp$ passbands along with linear fits over the range M8--L8 for other
passbands.

\end{abstract}

\section{Introduction}
\label{section.introduction}

The \G\ second data release \cite[hereafter \DR;][]{2018A&A...616A...1G} was
made on April 25th 2018 and contains parallaxes, proper motions, and magnitudes
for over one billion objects. The main astrometric observations use a
large optical passband called the $G$ band, and the completeness magnitude
goal of this mission in this band is 20.7\,mag \citep{2016A&A...595A...2G}. We
are interested in ultra-cool dwarfs (hereafter UCDs), defined as objects with
a spectral type later than M7. UCDs are intrinsically very faint in the
optical and, therefore, only limited numbers will be observable by \G. In
particular, we expect there to be around a 1000 L dwarfs and only a few T
dwarfs \citep{2002EAS.....2..199H, 2013A&A...550A..44S, 2014MmSAI..85..649S,
  2017MNRAS.469..401S}.

While this sample is relatively limited in numbers, the availability of all-sky
uniformly derived parallaxes provides a volume limited sample that is
very useful for a number of astrophysical problems. \G\ UCDs include objects
with masses that straddle the stellar--sub-stellar transition, and 
therefore help us define the observational boundary between hydrogen-burning
stars and degenerate brown dwarfs
\citep[e.g. see][]{2009AIPC.1094..102C, 2011ApJ...736...47B}. The final volume
limited sample will be used to model the stellar--sub-stellar mass
function \citep{2005ApJ...625..385A} and luminosity function
\citep{2007AJ....133..439C}, removing incompleteness and observational biases
(e.g. Malmquist, Eddington and Lutz-Kelker effects) that plague current
measurements of this fundamental observable \citep[e.g. see ][and references
  therein]{2012ApJ...753..156K, 2015MNRAS.449.3651M}.

The \G\ astrometry and photometry will provide robust measurements of
luminosity. The \G\ dataset will aid the modelling of the atmospheres of
low-mass objects by providing a cohort of new benchmark systems, such as
companions to main-sequence stars \citep{2017MNRAS.470.4885M,
  2018MNRAS.479.1332M} and members of young moving groups
\citep[e.g.][]{2015ApJ...798...73G}. L dwarfs are analogues for understanding
planetary atmospheres \citep{2016ApJS..225...10F} and, once we calibrate a
cooling curve \citep[e.g. by studying L dwarf companions to white
  dwarfs;][]{2011MNRAS.410..705D}, their ubiquity will make them promising
Galactic chronometers \citep{2010ARA&A..48..581S, 2009IAUS..258..317B}.

A first step in identifying \G\ L and T (hereafter LT) dwarfs was carried out
in %
\citet[][hereafter \SMA]{2017MNRAS.469..401S}, matching known LT dwarfs to
the first \G\ data release \citep{2016A&A...595A...2G}, which contained
accurate positions and $G$ magnitudes for 1.14 billion objects. This
cross-match resulted in 321 LT dwarfs with \G\ $G$ magnitudes
and positions. This catalogue makes up the cool part of the \G\ Ultra-cool
Dwarf Sample (hereafter GUCDS), which is being used as a training set in
Coordination Unit 8 of the \G\ Data Processing and Analysis Consortium
pipeline\footnote{\url{https://www.cosmos.esa.int/web/gaia/coordination-units}}.
In addition, \cite{2018A&A...616A..10G} cross-matched the input catalogue from
\SMA\ with \DR\ and external catalogues such as 2MASS
(\citealt{2006AJ....131.1163S}).  This exercise provided 601 LT dwarfs,
including 527 fully characterised objects. Here, we build on this legacy and
carry out a more comprehensive analysis.

In this paper we concentrate on the LT dwarfs that are in the \DR.  In
Section~\ref{section:catalogue} we describe the input catalogue of LT dwarfs
used to search the \DR, the cleaning carried out, and the production of the LT
part of the GUCDS catalogue. In Section~\ref{section:BinarySystems} we look at
LT dwarfs that are in binary systems with other objects in the \DR. In
Section~\ref{section:PhotometricExamination} we examine this catalogue in
absolute magnitude, colour, and spectroscopic space. In
the last section we give conclusions
and future plans.

\section{The comparison catalogues}
\label{section:catalogue}
\subsection{The \DR\ Selection}
Each of the 1332 million \DR\ sources with full astrometric solutions are the
result of individual five-parameter fits to their epoch positions. It is
inevitable that some of these fits produce physically nonsensical solutions
with large negative parallaxes being the most obvious examples. The solutions
with large positive parallaxes that appear to be nearby objects represent the
tail of the $10^9$ solutions distribution and is, in a relative sense,
significantly impacted by objects being scattered into that solution
space. Indeed, if one orders \DR\ by parallax, Proxima Centauri, the closest
object to the Sun, would be ranked \POSITIONOFPROX st. If we
consider objects with parallaxes greater than 200\,mas (i.e. distance $d <$
5\,pc), there are \NUMINFIVEPC\ of them in the \DR. However, only
\NUMWITHSIMBADINFIVEPC\ have a parallax in SIMBAD \citep{2000A&AS..143....9W}
that is greater than 200\,mas. There are also \NUMWITHSIMBADOUTFIVEPC\ of the
\NUMINFIVEPC\ objects that match to SIMBAD entries but have parallaxes or
photometric distances that place them at distances greater than 5\,pc. While
there is a remote possibility that some of the new objects with parallaxes
greater than 200\,mas in the \DR\ are really within 5\,pc, the majority, if
not all, of the remaining 754 solutions are incorrect.

In \citet[][their Appendix C]{2018A&A...616A...2L} they convincingly argued
that many of these bad solutions are due to mismatches of the
observations. They showed, as it would be expected if this is the dominant
reason, that the number of objects with large negative parallaxes is
approximately equal to the number of sources with large spurious positive
parallaxes. They also provided a number of quality cuts that would reduce the
contamination at a small cost to the identification of real objects. However,
as our final goal is to make a complete census of all UCDs in the \G\ dataset,
we want our training set to include also objects with low quality astrometry
so we do not apply those cuts.  In addition, the \DR\ is missing astrometric
solutions for prominent nearby bright LT dwarfs, e.g. $WISE$ J1049-5319A,
\citep{2013APJ...767L...1L} and $\epsilon$~Indi~B~ab
\citep{2003A&A...398L..29S}, probably because these are binary systems with
large orbital motions and their solutions did not meet the quality thresholds
for inclusion in the \DR.

Since the majority of large parallaxes are unreliable and some of the nearest
objects are missing, it is premature to attempt to find all UCDs to the
\G\ magnitude limit and, therefore, we concentrate on developing criteria for
a robust selection procedure in the future. The first step in developing such
criteria is the identification of known UCDs that we can use as a training
set. In \SMA\ we showed that the most distant single L0 that we expected to see in
\G\ is at 80\,pc. There are unresolved binary L dwarf systems outside the
100\,pc limit that have a combined magnitude greater than the
\DR\ limit. There are also very young L dwarfs that have very bright intrinsic
magnitudes for their spectral type and these may enter the \DR\ even though they
are at a distance greater than 100\,pc. For example some of the L dwarfs
identified in the Upper Scorpius OB association \citep[see][ and reference
  therein]{2008MNRAS.383.1385L} at a distance of 145$\pm$2\,pc
\citep{1999AJ....117..354D} with an age of 5\,Myr \citep{1999AJ....117.2381P}
have predicted \G\ apparent magnitudes $G < 20$\,mag from \SMA. However, the
vast majority of LT dwarfs seen by \G\ are within 80\,pc, so we start by
selecting all objects from the \DR\ with a parallax greater than 10\,mas, e.g.
a distance limit of 100\,pc, which results in 700,055 sources.

\subsection{The M, L, T or Y catalogue}
\begin{table*}
 \caption{Input catalog entries with multiple \DR\ matches within
   $20$\arcsec. Where both objects are in the input catalog of UCDs we include
   their names. Objects with ``Bin'' in the first column are found to be
   physical binaries based on the test discussed in
   Section~\ref{BinarySystems}.}
 \label{multimatch}
 \begin{tabular}{llcccccccc}   
\hline                                                
Short & Discovery Name & Offset      & $G$ & $G-G_{\rm est}$  & $\varpi $ &
$\mu_{\rm tot}$  &$ \theta_{\mu} $ \\
name     & and  \dt{Source\_ID}     & \arcsec\    & mag &  mag        &  mas    & mas\,yr$^{-1}$ &  \degr \\
\hline                                      
          J0004-4044  &                     GJ 1001 B  &     0.8 &   18.353 &  -0.4 &    82.1 $\pm$   0.4 &  1641.6 &   155.9\\
\smallskip
        ~~~~~~~~~~Bin &            4996141155411983744 &    18.4 &   11.500 &  -7.2 &    81.2 $\pm$   0.1 &  1650.7 &   155.8\\

          J0235-2331  &                     GJ 1048 B  &     0.1 &   18.598 &   0.6 &    46.6 $\pm$   0.3 &    97.0 &    77.5\\
\smallskip                      &            5125414998097353600 &    12.1 &    7.987 & -10.1 &    47.1 $\pm$   0.0 &    84.5 &    80.4\\
          J0858+2710  &        2MASS 08583693+2710518  &     0.1 &   19.926 &   0.3 &    18.9 $\pm$   1.3 &   221.4 &   155.9\\
\smallskip        ~~~~~~~~~~Bin &             692611481331037952 &    15.2 &   15.067 &  -4.6 &    17.9 $\pm$   0.1 &   215.0 &   156.8\\

          J1004+5022  &                     G 196-3 B  &     0.2 &   20.170 &   0.3 &    44.4 $\pm$   0.8 &   250.0 &   213.5\\
\smallskip        ~~~~~~~~~~Bin &             824017070904063104 &    15.9 &   10.612 &  -9.3 &    45.9 $\pm$   0.0 &   246.8 &   214.9\\

          J1004-3335  &         2MASSWJ1004392-333518  &     0.3 &   19.615 &   0.3 &    53.3 $\pm$   0.6 &   495.1 &   135.7\\
\smallskip        ~~~~~~~~~~Bin &            5458784415381054464 &    12.0 &   12.908 &  -6.5 &    53.5 $\pm$   0.1 &   488.8 &   135.0\\

          J1047+4046  &                     LP213-067  &     4.4 &   15.183 &  -1.3 &    40.1 $\pm$   0.1 &   299.9 &   263.7\\
\smallskip          J1047+4047  &                     LP213-068  &     4.5 &   16.931 &  -0.7 &    38.9 $\pm$   0.5 &   303.3 &   263.5\\

          J1202+4204  &        2MASS 12025009+4204531  &     0.2 &   19.321 &  -0.5 &    31.5 $\pm$   0.4 &   366.6 &   217.5\\
\smallskip        ~~~~~~~~~~Bin &            1537249785437526784 &     7.8 &   16.430 &  -3.4 &    31.6 $\pm$   0.1 &   368.9 &   218.3\\
          J1219+0154  &      ULAS J121932.54+015433.0  &     0.1 &   19.792 &  -0.6 &    18.9 $\pm$   0.6 &   114.9 &   229.9\\
\smallskip        ~~~~~~~~~~Bin &            3700975728440669184 &    10.9 &   13.441 &  -6.9 &    19.8 $\pm$   0.1 &   115.2 &   230.4\\
          J1245+0156  &      ULAS J124531.54+015630.9  &     0.1 &   20.612 &  -0.5 &    13.5 $\pm$   1.2 &    76.0 &   234.7\\
\smallskip        ~~~~~~~~~~Bin &            3702489721592680832 &     8.2 &   12.860 &  -8.3 &    13.2 $\pm$   0.0 &    75.6 &   235.1\\
          J1304+0907  &        2MASS 13043318+0907070  &     0.1 &   20.173 &  -0.4 &    18.2 $\pm$   0.8 &   134.8 &   278.6\\
\smallskip        ~~~~~~~~~~Bin &            3734192764990097408 &     7.6 &   15.160 &  -5.4 &    17.8 $\pm$   0.1 &   134.3 &   277.9\\

          J1442+6603A &                    G 239-25 A  &     0.2 &    9.851 &  -2.3 &    91.5 $\pm$   0.0 &   301.6 &   262.6\\
\smallskip          J1442+6603  &                    G 239-25 B  &     0.2 &   15.302 &  -1.4 &    91.7 $\pm$   0.2 &   338.5 &   274.3\\
          J1520-4422  &              WDS J15200-4423A  &     0.4 &   18.293 &  -0.3 &    54.5 $\pm$   0.2 &   736.7 &   238.6\\
\smallskip          J1520-4422B &              WDS J15200-4423B  &     0.4 &   19.817 &   1.0 &    53.7 $\pm$   0.6 &   753.4 &   238.6\\
          J1540+0102  &      ULAS J154005.10+010208.7  &     0.0 &   19.851 &  -0.7 &    14.8 $\pm$   0.6 &    51.7 &   253.1\\
  \smallskip                    &            4416887712294719104 &    13.6 &   14.863 &  -5.6 &    17.0 $\pm$   0.7 &    50.9 &   267.0\\

          J1711+4028  &                    G 203-50 B  &     5.5 &   20.232 &  -0.4 &    47.4 $\pm$   0.7 &   263.5 &    72.4\\
\smallskip        ~~~~~~~~~~Bin &            1341903196663707904 &     8.7 &   14.233 &  -6.4 &    47.1 $\pm$   0.1 &   265.5 &    72.0\\

          J2200-3038A &   DENIS-PJ220002.05-303832.9A  &     3.7 &   18.437 &  -0.3 &    25.4 $\pm$   0.4 &   247.2 &   104.9\\
\smallskip          J2200-3038B &   DENIS-PJ220002.05-303832.9B  &     0.1 &   19.042 &  -0.6 &    25.3 $\pm$   0.5 &   253.7 &   105.6\\

          J2308+0629  &      ULAS J230818.73+062951.4  &     0.1 &   18.059 &  -0.7 &    24.7 $\pm$   0.3 &   118.5 &   162.3\\
\smallskip        ~~~~~~~~~~Bin &            2665079816223169664 &     3.8 &   13.467 &  -5.3 &    24.1 $\pm$   0.1 &   119.8 &   160.5\\

          J2322-6151  &        2MASS 23225299-6151275  &     0.0 &   20.682 &   0.3 &    23.2 $\pm$   1.0 &   114.6 &   135.7\\
       ~~~~~~~~~~Bin &            6487249243899899904 &    16.6 &   14.902 &  -5.5 &    23.6 $\pm$   0.1 &   110.3 &   135.2\\

\hline
\end{tabular}
\end{table*}

The initial list of known UCDs was the input catalogue from \SMA\ of 1885
objects with M, L, T or Y spectral classification. To this we added the
photometrically-identified LT dwarfs from \cite{2016A&A...589A..49S} and a few
other recent discoveries \cite[e.g.][]{2017MNRAS.470.4885M,
  2018RNAAS...2...33S, 2018MNRAS.474.1826S}.
The current list contains \NKNOWNLT\ UCDs ranging from M8 to Y2 dwarfs. The M
dwarfs were retained to facilitate differentiation of the spectral types in
the magnitude and colour space, as some objects are classed as M in optical
spectra and L in infrared spectra and vice-versa.  Since this input catalogue
is dominated by L and T dwarfs we refer to it as the LT catalogue  (hereafter
LTC). 

For all objects we have collected photometry from the Two Micron All Sky
Survey \citep[2MASS;\,][]{2006AJ....131.1163S}, the Panoramic Survey Telescope
and Rapid Response System release 1 \citep[PS1;\,][]{2016arXiv161205560C}, and
the {\it Wide-field Infrared Survey Explorer}
extension \citep[AllWISE;\,][]{2010AJ....140.1868W}, providing up-to 11
homogeneous magnitudes in passbands ranging from Gunn $g$ to $WISE$ $W3$. The
$WISE$ $W4$ band was not included as the number of objects with reliable
magnitudes were very low.

We started by searching for any objects in the \DR\ release that had a
parallax larger than 10\,mas and was within $20$\arcsec\ of the LTC
entry at the \DR\ epoch. We choose $20$\arcsec\ as not all entries have
published proper motions,  the
epoch difference can be up to 20 years, and typical proper motions are
500--1000\masyr. This resulted in \NLARGEMATCH\ entries from \DR. For each
entry we then propagated the \DR\ position to the epoch of the LTC positions in
the input catalogue using the \DR\ proper motions.

\subsection{Treatment of duplicate matches}
\label{section:duplicatematches}

Some multiple LTC dwarfs matched to the same \DR\ source,
e.g. J1416+1348A/J1416+1348B to 1227133699053734528 and
J1207-3932A/J1207-3932B (TWA 27 A and B) to 3459372646830687104. These are
known binary systems where both the primary and secondary are in our LTC and,
the primary is observed by \G\ but the secondary is not. This maybe because the
secondary is faint, e.g. J1416+1348B has an estimated $G=23.9$, or the primary
and secondary are very close and are not resolved in the \DR\, e.g TWA 27A/B
have a separation of 0.7\arcsec \citep{2004A&A...425L..29C}.  For these
multiple LTC entry matches we assumed that the correct match was the brightest
of the binary system.

In general the closest of multiple matches in the \DR\ at the epoch of the LTC
position is the correct one, but this may not be always the case. 
Using only unique matches we calibrated robust linear relations between
the \DR\ $G$ magnitude and the optical spectral types for each LTC entry with
external optical photometry, and the NIR spectral types with external
NIR photometry. Using these relations, we estimated an average 
magnitude $G_{\rm est}$ for all objects in the LTC (see \SMA). There were
\NMULTIMATCH\ LTC entries with more than one \DR\ object within
$20$\arcsec.  In Table\,\ref{multimatch}
we report  for these objects the source identification number, \dt{Source\_ID}, of the
matched \DR\ entries, the offset on the sky of the LTC
position (generally 2MASS coordinates) and the
\DR\ entry, $G$ magnitude, $G-G_{\rm est}$ difference, parallax in milli-arcseconds (hereafter
mas), total proper motion in~\masyr\ and position angle of the proper
motion in degrees.

In all cases the nearest positional match also has the smallest $G-G_{\rm
  est}$, consistent with being the LT dwarf. Extra objects within $20$\arcsec\
are in most cases the other component in known physical binary systems, as is evident 
when the ``Discovery Name'' indicates the LT dwarf is the B
component in a known system.  To all these combinations we applied the binary
test described in Section~\ref{BinarySystems}, and if a pair satisfies
it - i.e. we consider the pair to be a physical system - we label it `Bin' in
the first column.

There are three matches that did not pass our binarity test: J0235-2331,
J1442+6603A, and J1540+0102. Both J0235-2331 (GJ 1048 B) and J1442+6603
(G239-25B) are in known binary systems where the primary has been correctly
identified in Table\,\ref{multimatch}. These represent a failure of our
binarity test; reasons for this could be that the orbital motion is
significant so the proper motions are not within 10\%, or simply bad
solutions; we discuss this in Section~\ref{BinarySystems}. For J1540+0102, the
nearby object 4416887712294719104 has a parallax that differs by more than
3$\sigma$, but it is at the limit and has consistent proper motions, so it
warrants further consideration. Most binary systems are already noted in the
literature except J1219+0154 and J2308+0629, which were first published in
\cite{2016A&A...589A..49S} and photometrically classified as single L
dwarfs. Since this is quite a recent study, the entries have not received a
significant amount of follow up, so new candidate binary system
discoveries are not unexpected. Systems of particular interest are discussed in
Section~\ref{BinarySystems}.

\subsection{Cleaning of matched objects}
\begin{table*}
 \caption{{\it Top:} intrinsically bright ($M_{G} < 14.0$ mag) objects matched
   to LT objects sorted by absolute magnitude. {\it Bottom:} matches with
  offset $> 5.0$\arcsec\ sorted by separation.}
 \label{bright}
 \begin{tabular}{lllcccc}   
\hline                                                
Short    & Discovery  & Source\_ID  & Offset      & $G$~ &  $G_{est}$ & $M_G$ \\
name     & name       &             & \arcsec\    & mag &  mag      &  mag  \\  
\hline                                      
 J1207-3932A &                 TWA 27 A  &  3459372646830687104 &   0.02 &   17.408 &   17.450 &   13.363 \\
 J1610-0040  &           LSRJ 1610-0040  &  4406489184157821952 &   0.67 &   16.595 &   17.066 &   13.917 \\
 J0133-6314  &         SSSPM J0134-6315  &  4712132354155559040 &   0.05 &   18.206 &   19.652 &   13.919 \\

\hline
         \noalign{\smallskip}
 J1711+4028  &               G 203-50 B  &  1341903196662424320 &   5.50 &   20.232 &   20.411 &   18.613 \\
 J2250+0808  &                 BRLT 317  &  2713153831843361920 &   5.98 &   20.642 &   20.410 &   17.456 \\

\hline
\end{tabular}
\end{table*}

The majority (93\%) of objects in the \DR\ with $\varpi > 10$\,mas have $
\varpi/\sigma_{\varpi} > 5.$ Since we are considering each object
individually, for our selection purposes it is sufficient to use a simple
distance given by the inverse of the parallax, $1/\varpi$
\citep{2015PASP..127..994B}.  Using this distance, we calculated an absolute
magnitude in the $G$ band, $M_G$. Considering the bulk of L0 dwarfs, we found
that a conservative absolute magnitude limit for this spectral type is
$M_G$=14.0\,mag. Among the remaining \NMATCHED\ matched objects, only
\NBRIGHT\ are brighter than this magnitude which we visually inspected. Often
they were incorrect matches where the LTC entry is a companion in a binary
system that was too faint for \G, so instead we matched to the bright
component, e.g. the T7 dwarf GJ\,229B was matched to its M1\,V primary
GJ\,229A. However, some were just close unrelated stars, e.g.  J1119+0021 is a
T4.5 that was too faint for \G\ but matched to the unrelated object
UCAC4\,452-049871. On individual inspection of the \NBRIGHT\ sources, only
J1207-3932A, J1610-0040 and J0133-6314 appeared to be correct matches to
late-type M dwarfs, and all other matches were removed. J1047+4046 was also
probably correctly matched, but to a M6.5, so we did not retain it as it is
outside our M7 limit. The details of these three objects are included in
Table~\ref{bright}.

This absolute magnitude test was not possible for \SMA\, because there were no
parallaxes, required to calculate distance moduli. In
\cite{2018A&A...616A..10G} they used the \SMA\ input catalogue and did not
apply this cleaning, so the very bright ``LT objects'' on the main sequence of
their Figure 9a and the white dwarf track of their Figure 9c are matches of
the bright primary star to the fainter LT companion in the \SMA\ input
catalogue.

After removing the bright objects we found \NLARGEDIST\ objects with an offset
from the predicted position and the \DR\ position larger than $5$\arcsec. We
retained the first two (J1711+4028, and J2250+0808), listed in
Table~\ref{bright}, as both the magnitude difference $G-G_{\rm est}$ was not
very large, and in a visual inspection of the region there did not appear to
be any other nearby objects. For the other three objects (J1108+1535,
J1928+2356, and J1456-2747) they have a large offset from the LTC
predicted position (19.4, 19.6 and 19.9\arcsec respectively) and a large
$G-G_{\rm est}$\,mag difference (-0.8, -2.3 and -2.8\,mag,
respectively). We conclude that these are objects undetected by \G\ that have
been matched to a nearby
unrelated star. The target J1928+2356 has a $G_{\rm est}=20.182$\,mag, 
nominally within the \G\ magnitude limit and may appear in later releases, but the other
two both have  $G_{\rm est}>21.0$\,mag, so they will probably not be detected.

\subsection{The GUCDS catalogue}
\begin{figure}
\begin{center}
\includegraphics[scale=0.45]{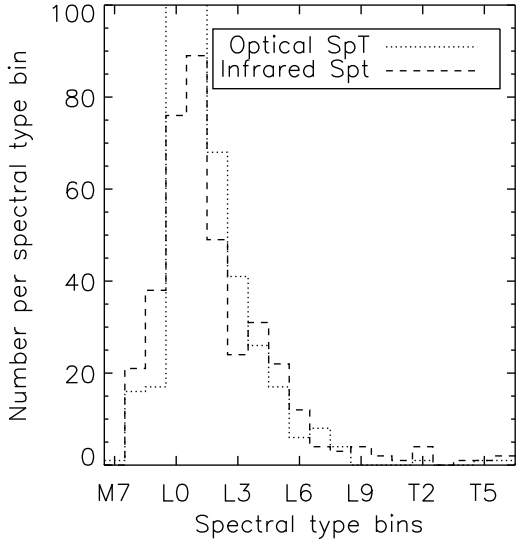}
\includegraphics[scale=0.45]{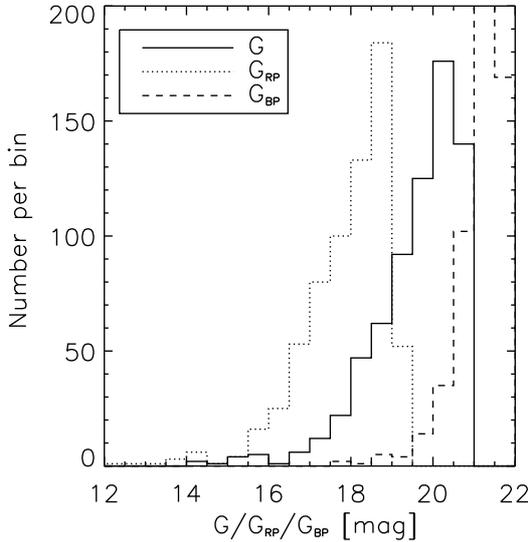}
\caption{{\it Top}: distribution of optical and infrared spectral types in the
  \Gcat. Optical L0 and L1 bins have been truncated (they contain 202 and 105
  objects, respectively). {\it Bottom}: distribution of \G\ magnitudes in the
  \Gcat. }
\label{SptDist}
\end{center}
\end{figure}

\begin{table*}
 \caption{Content of the \Gcat\ with J1807+5015 as an example.}
 \label{allcat}
 \begin{tabular}{lllll}   
\hline                                                
Parameter & Format & Unit & Comment & Example  \\
\hline                                      
              SHORTNAME &    a12 & ...    &              Short name used in text of paper      &  J1807+5015 \\
                     RA &  f13.9 & deg    &              Right ascension (eq. J2000, ep. 2015) & 271.816572024\\
                    DEC &  f13.9 & deg    &              Declination (eq. J2000, ep. 2015)     &  50.258197767\\
          DISCOVERYNAME &    a25 & ...    &              Common discovery name                 &   2MASSI J1807159+501531 \\
       DISCOVERYREFNAME &    a19 & ...    &              Discovery reference                   & 2003AJ....126.2421C\\
              SOURCE\_ID &    i20 & ...    &              Gaia DR2 source ID                    &  2123161836615550848\\
             DISTARCSEC &   f6.2 & ...    &              Distance DR2 to catalog position      &   0.14\\
       MULTIPLEFLAGNAME &    a10 & ...    &              Multiple code VB/UR/MG                &        ...\\
    MULTIPLEFLAGREFNAME &    a19 & ...    &              Multiple code reference BibCodes      &                 ...\\
             SPTOPTNAME &    a10 & ...    &              Optical Spectral type                 &      L1.5 \\
          SPTOPTREFNAME &    a19 & ...    &              Optical Spectral type BibCode         & 2003AJ....126.2421C\\
             SPTNIRNAME &    a10 & ...    &              Near infrared Spectral type           &        L1 \\
          SPTNIRREFNAME &    a19 & ...    &              Near infrared Spectral type BibCode   & 2003IAUS..211..197W\\
             SPTPHONAME &    a10 & ...    &              Photometric Spectral type             &       ... \\
          SPTPHOREFNAME &    a19 & ...    &              Photometric Spectral type BibCode     &                ... \\
           LIT\_PARALLAX &  f10.3 & mas    &              Published parallax                    &     77.250\\
     LIT\_PARALLAX\_ERROR &  f10.3 & mas    &              Published parallax error              &      1.480\\
    LIT\_PARALLAXREFNAME &    a19 & ...    &              Published parallax  BibCode           & 2014PASP..126...15W\\
                 TMASSJ &  f10.3 & mag    &              2MASS J band magnitude                &     12.934\\
              TMASSJERR &  f10.3 & mag    &              2MASS J band magnitude error          &      0.024\\
                 TMASSH &  f10.3 & mag    &              2MASS H band magnitude                &     12.127\\
              TMASSHERR &  f10.3 & mag    &              2MASS H band magnitude error          &      0.031\\
                 TMASSK &  f10.3 & mag    &              2MASS K band magnitude                &     11.602\\
              TMASSKERR &  f10.3 & mag    &              2MASS K band magnitude error          &      0.025\\
                 WISEW1 &  f10.3 & mag    &              ALLWISE W1 Band magntiude             &     11.246\\
              WISEW1ERR &  f10.3 & mag    &              ALLWISE W1 Band magntiude error       &      0.023\\
                 WISEW2 &  f10.3 & mag    &              ALLWISE W2 Band magntiude             &     10.971\\
              WISEW2ERR &  f10.3 & mag    &              ALLWISE W2 Band magntiude error       &      0.021\\
                 WISEW3 &  f10.3 & mag    &              ALLWISE W3 Band magntiude             &     10.505\\
              WISEW3ERR &  f10.3 & mag    &              ALLWISE W3 Band magntiude error       &      0.056\\
                  GUNNG &  f10.3 & mag    &              PANSTARRS G Band magntiude            &     21.955\\
               GUNNGERR &  f10.3 & mag    &              PANSTARRS G Band magntiude error      &      0.061\\
                  GUNNR &  f10.3 & mag    &              PANSTARRS R Band magntiude            &     19.748\\
               GUNNRERR &  f10.3 & mag    &              PANSTARRS R Band magntiude error      &      0.013\\
                  GUNNI &  f10.3 & mag    &              PANSTARRS I Band magntiude            &     17.375\\
               GUNNIERR &  f10.3 & mag    &              PANSTARRS I Band magntiude error      &      0.003\\
                  GUNNZ &  f10.3 & mag    &              PANSTARRS Z Band magntiude            &     15.925\\
               GUNNZERR &  f10.3 & mag    &              PANSTARRS Z Band magntiude error      &      0.005\\
                  GUNNY &  f10.3 & mag    &              PANSTARRS Y Band magntiude            &     14.936\\
               GUNNYERR &  f10.3 & mag    &              PANSTARRS Y Band magntiude error      &      0.006\\
        PHOT\_G\_MEAN\_MAG &  f10.3 & mag    &              Gaia DR2  G Band magntiude            &     17.807\\
  PHOT\_G\_MEAN\_MAG\_ERROR &  f10.3 & mag    &              Gaia DR1  G Band magntiude error      &      0.002\\
       PHOT\_G\_MEAN\_FLUX &  f10.1 & ...    &              Gaia DR2  G Band flux                 &     1420.5\\
 PHOT\_G\_MEAN\_FLUX\_ERROR &   f8.1 & ...    &              Gaia DR1  G Band flux error           &      2.2\\
       PHOT\_BP\_MEAN\_MAG &  f10.3 & mag    &              Gaia DR2 BP Band magntiude            &     20.931\\
 PHOT\_BP\_MEAN\_MAG\_ERROR &  f10.3 & mag    &              Gaia DR1 BP Band magntiude error      &      0.137\\
      PHOT\_BP\_MEAN\_FLUX &  f10.1 & ...    &              Gaia DR2 BP Band flux                 &       58.6\\
PHOT\_BP\_MEAN\_FLUX\_ERROR &   f8.1 & ...    &              Gaia DR1 BP Band flux error           &      7.4\\
       PHOT\_RP\_MEAN\_MAG &  f10.3 & mag    &              Gaia DR2 RP Band magntiude            &     16.193\\
 PHOT\_RP\_MEAN\_MAG\_ERROR &  f10.3 & mag    &              Gaia DR1 RP Band magntiude error      &      0.006\\
      PHOT\_RP\_MEAN\_FLUX &  f10.1 & ...    &              Gaia DR2 RP Band flux                 &     2676.0\\
PHOT\_RP\_MEAN\_FLUX\_ERROR &   f8.1 & ...    &              Gaia DR1 RP Band flux error           &     15.1\\
               GAIAGEST &  f10.3 & mag    &              Estimated  DR2  G  from SpT           &     17.978\\
               PARALLAX &   f8.2 & mas    &              Gaia DR2 parallax                     &    68.33\\
         PARALLAX\_ERROR &   f5.2 & mas    &              Gaia DR2 parallax error               &  0.13\\
                   PMRA &   f8.2 & mas/yr &              Gaia DR2 Proper motion in RA          &    24.49\\
             PMRA\_ERROR &   f5.2 & mas/yr &              Gaia DR2 RA proper motion error       &  0.25\\
                  PMDEC &   f8.2 & mas/yr &              Gaia DR2 Proper motion in  Dec        &  -136.91\\
            PMDEC\_ERROR &   f5.2 & mas/yr &              Gaia DR2 Dec proper motion error      &  0.27\\

\hline
\end{tabular}
\end{table*}

After cleaning the initial match, our final catalogue is made up of
\NINCAT\ objects in the spectral range M8 to T6 with \G\ astrometry.  In the
top panel of Figure~\ref{SptDist} we show the distribution of the
\NSPTOPT\ objects with optical spectral classification and the \NSPTNIR\ with
infrared spectral classification. There were \NUNRESOLVED\ unresolved systems
where we assumed, for the distributions in Figure~\ref{SptDist}, the earliest of the two spectral
types. For example, J0320-0446 has an infrared spectral type of ``M8.5 + T5:''
\citep{2008APJ...689L..53B}; for the distribution we assumed a spectral
type of M8.5. There were also \NSPTPHO\ objects with just a photometric
spectral type from \cite{2016A&A...589A..49S} that are not included in 
these figures.

In the lower panel of Figure~\ref{SptDist} we show the $G$, $\grp$, and $\gbp$
magnitude distributions.  All 695 entries have a $G$ magnitude (as well
as a proper motion and a parallax), as this is a requirement for inclusion in
the \DR, and \NRP\ UCDs have $\gbp$ and $\grp$ magnitudes. In
Table~\ref{allcat} we list the astrometry, spectroscopy, photometry and other
parameters for the catalogue that are used in the following sections. The full
catalogue is available online here and we will refer to it as the \Gcat.

\subsection{Comparison of parallaxes with published results}
\begin{figure}
\label{New_Published_pi}
\begin{center}
\includegraphics[scale=0.45]{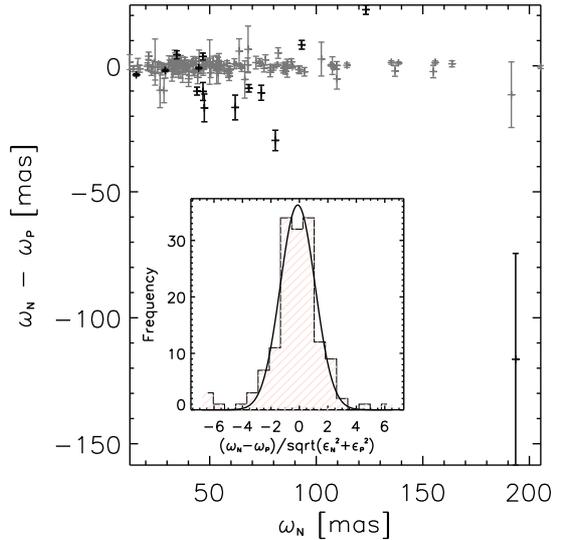}
\caption{Differences between \DR\ and published parallaxes vs. \DR\ parallaxes.  Error bars are
  combined \DR\ and published uncertainties plotted in grey. The large outlier is J1506+7027
  discussed in the text. In black, all object with parallaxes differing by
  more than  2.5 times the combined uncertainties (listed in
  Table~\ref{piComparison}). The insert is a plot of the distribution of the
  ratio of parallax differences to combined uncertainties as shown in Equation~\ref{ratio}}
\end{center}
\end{figure}

\begin{table}
\begin{center}
 \caption{Objects with published parallax estimates differing by more than
   2.5$\sigma$ with the \DR\ parallax.}
 \label{piComparison}
 \begin{tabular}{lcc}   
\hline                                                
         \noalign{\smallskip}
Short    & \G\ $\varpi$  & Published $\varpi$  \\
Name     & mas           &    mas                     \\  
\hline                                      
         \noalign{\smallskip}
J0439-2353  &   80.79 $\pm$  0.51 &  110.40 $\pm$  4.00$^{1}$\\
J0445-3048  &   61.97 $\pm$  0.18 &   78.50 $\pm$  4.90$^{1}$\\
J0615-0100  &   44.80 $\pm$  0.33 &   45.70 $\pm$  0.11$^{2}$\\
J0805+4812  &   46.78 $\pm$  0.96 &   43.10 $\pm$  1.00$^{3}$\\
J1017+1308  &   34.56 $\pm$  0.82 &   30.20 $\pm$  1.40$^{3}$\\
J1155-3727  &   84.57 $\pm$  0.19 &  104.38 $\pm$  4.69$^{1}$\\
J1207-3932A &   15.52 $\pm$  0.16 &   19.10 $\pm$  0.40$^{4}$\\
J1254-0122  &   74.18 $\pm$  2.31 &   84.90 $\pm$  1.90$^{5}$\\
J1359-4034  &   47.51 $\pm$  0.27 &   64.18 $\pm$  5.48$^{1}$\\
J1454-6604  &   93.22 $\pm$  0.30 &   84.88 $\pm$  1.71$^{6}$\\
J1506+7027  &  193.55 $\pm$  0.94 &  310.00 $\pm$ 42.00$^{7}$\\
J1610-0040  &   29.14 $\pm$  0.37 &   31.02 $\pm$  0.26$^{8}$\\
J1717+6526  &   46.86 $\pm$  0.62 &   57.05 $\pm$  3.51$^{9}$\\
J1731+2721  &   83.74 $\pm$  0.12 &  113.80 $\pm$  7.00$^{10}$\\
J1807+5015  &   68.33 $\pm$  0.13 &   77.25 $\pm$  1.48$^{9}$\\
J2148+4003  &  123.28 $\pm$  0.46 &  101.01 $\pm$  1.78$^{11}$\\
\noalign{\smallskip} \hline
\end{tabular}
\end{center}
Discovery references - 
1: \cite{2012ApJ...752...56F},  2: \cite{2014A&A...565A..20S},   
3: \cite{2012ApJS..201...19D},  4: \cite{2008A&A...477L...1D},   
5: \cite{2002AJ....124.1170D},  6: \cite{2014AJ....147...94D},   
7: \cite{2013ApJ...762..119M},  8: \cite{2008ApJ...686..548D},   
9: \cite{2014PASP..126...15W},  10: \cite{2014ApJ...784..156D},  
11: \cite{2016ApJ...833...96L}                                   
\end{table}

In the \Gcat\ \NWITHPI\ entries have previously published parallaxes. In
Figure~\ref{New_Published_pi} we plot the \DR\ versus the published
values. In Table~\ref{piComparison} we have listed all objects with \DR\ and
published values that differ by more than 2.5 times the combined uncertainties.
There is only one significant outlier, J1506+7027, which had a parallax
estimated in \cite{2013ApJ...762..119M} of 310$\pm$42\,mas using eight epochs
over two years from a combination of $WISE$, WIRC and $Spitzer$ compared to
the \DR\ value of 193.5$\pm$0.9\,mas. The photometric parallax for this object
would be 187\,mas based on the apparent magnitude-spectral type of the
\cite{2012ApJS..201...19D} calibration, consistent with the \DR\ value. It is
very difficult to successfully combine observations from different instruments
in small field astrometry, and the \DR\ solution does not give any indication
of problem. Therefore we adopt the \G\ value.

The \DR\ parallaxes have a median uncertainty of 0.4\,mas while the published
parallaxes have a median uncertainty of 1.5\,mas. For the objects with published
parallaxes we calculated the ratio
\begin{equation}
\label{ratio}
r = \frac{\varpi_{N} - \varpi_{P}}{\sqrt{\sigma_{N}^2 + \sigma_{p}^2 }},
\end{equation}
where $\varpi$ is the parallax, $\sigma$ the quoted uncertainties, and the
subscripts $N$ and $P$ represent the new and published values,
respectively. If the measures were unbiased and the uncertainties correct we
would expect this ratio to follow a Gaussian distribution with a mean of zero
and a standard deviation of unity. For the
\NWITHPI\ common objects, after 3$\sigma$ clipping, the mean is -0.02 and the
standard deviation is 1.3.  Applying the t-test at the 95\% level we find that
the mean is not consistent with zero, i.e. P(t)=0.048, while applying the
F-test we find that the $\sigma$ is significantly different from one,
e.g. P(F)=$2\times 10^{-6}$.  Since the $\sigma$ of the ratio is greater than
unity, the implication is that the uncertainties are underestimated. To reconcile
the differences, the uncertainties of the published values would have to be
increased by $\sim$120\%, or those of \G\ by $\sim$800\%.  However, as is
evident in Table~\ref{piComparison}, the source of published parallaxes is
very heterogeneous, and the calculation of the errors are functions of the
different program reduction routines. Hence to obtain applicable corrections
the sample should be split into the contributing programs and then
individually assessed.  The \DR\ will enable a characterisation of the
uncertainties of the different small field programs, and the \G\ parallaxes of
the anonymous field stars used in the programs allows a precise estimate
of the correction from relative to absolute parallax, which is one of the most
unreliable steps in small field astrometry. In this way \G\ will contribute to
an improvement in the determination and characterisation of parallaxes for
objects that are fainter than its magnitude limit.

\section{BINARY SYSTEMS}
\label{section:BinarySystems}
\begin{table*}
 \caption{New candidate binary systems containing LT dwarfs, identified in \DR.}
 \label{BinarySystems}
 \begin{tabular}{lccclcccc}   
\hline                                                
Discovery Name & $\rho$    & RA, Dec   &  Spec. & $G$ & $\varpi$ & $\mu_{tot}$  &$ \theta_{\mu} $  \\
               &  \arcsec\ & \degr\   &  Type  & mag &  mas                &  mas\,yr$^{-1}$  & \degr \\  
\hline                                      
2MASS J01415823-4633574  & \multirow{2}{*}{2377.2} & 25.4933685,-46.5661305 & L2.0   &  20.02 &    27.4 $\pm$   0.5 &   124.7 &   111.9 \\%
4954453580066220800      &                         & 26.1336010,-46.0756886 & \ldots &  15.67 &    25.9 $\pm$   0.1 &   117.3 &   111.3 \\%
\hline
2MASS J02235464-5815067  &        & 35.9785858,-58.2519130 & L1.5 &  20.22 &    24.4 $\pm$   0.6 &   105.6 &    99.5 \\%
UCAC4 159-002053         & 1532.6 & 35.2149151,-58.3948241 & M3   &  12.57 &    22.7 $\pm$   0.0 &    97.2 &    99.7 \\%
2MASS J02251947-5837295  & 1499.0 & 36.3319895,-58.6249554 & M9   &  18.41 &    24.3 $\pm$   0.2 &   102.0 &    99.1 \\%
\hline
2MASSI J0518461-275645   &  \multirow{2}{*}{1007.2} & 79.6925197,-27.9460523 & L1.0   &  20.48 &    17.3 $\pm$   0.8 &    32.6 &    98.7 \\%
2954995674982867968      &                          & 79.8573963,-28.1850235 & \ldots &  15.11 &    17.6 $\pm$   0.1 &    32.6 &    98.9 \\%
\hline
2MASS J08430796+3141297  & \multirow{2}{*}{819.5} & 130.7828536, 31.6913490 & L2.5   &  20.91 &    14.8 $\pm$   2.3 &    67.9 &   230.3 \\%
709905940243414400       &                        & 130.6127152, 31.8671235 & \ldots &  17.47 &    10.2 $\pm$   0.2 &    73.1 &   235.1 \\%
\hline
2MASS J09073765+4509359  & \multirow{2}{*}{301.1} & 136.9073579, 45.1597676 & M9.0   &  18.99 &    26.3 $\pm$   0.4 &    76.8 &   118.1 \\%
TYC 3424-215-1           &                        & 137.0239116, 45.1753675 & \ldots &   9.22 &    27.0 $\pm$   0.1 &    80.2 &   123.5 \\%
\hline
2MASS J09175035+2944455  & \multirow{2}{*}{1684.7} & 139.4595607, 29.7456267 & L0.0   &  20.74 &    18.5 $\pm$   2.4 &    81.2 &   215.9 \\%
698766581783119872       &                         & 139.8773149, 29.4505981 & \ldots &  17.65 &    12.8 $\pm$   0.3 &    74.4 &   227.6 \\%
\hline
2MASS J11414410+4116568  & \multirow{2}{*}{163.5} & 175.4341457, 41.2822985 & L0.0 &   20.36 &    13.2 $\pm$   1.1 &    60.1 &   133.9 \\%
HD101620                 &                        & 175.4433423, 41.2374147 & F5   &    6.79 &    12.7 $\pm$   0.0 &    58.7 &   130.9 \\%
\hline
SDSS J124514.95+120442.0  & \multirow{2}{*}{96.4} & 191.3122876, 12.0781604 & L1.0 &  20.98 &    12.3 $\pm$   3.0 &    54.8 &   191.1 \\%
SDSS J124520.60+120531.3  &                       & 191.3358362, 12.0918479 & DA   &  18.29 &    12.2 $\pm$   0.3 &    54.8 &   186.9 \\%
\hline
ULAS J124531.54+015630.9  & \multirow{2}{*}{8.2}  & 191.3813059,  1.9418705 & \ldots &   20.61 &    13.5 $\pm$   1.2 &    76.0 &   234.7 \\%
3702489721592680832       &                       & 191.3791501,  1.9411384 & \ldots &   12.86 &    13.2 $\pm$   0.0 &    75.6 &   235.1 \\%
\hline
WDS J15200-4423A  & \multirow{2}{*}{1.0} & 230.0053261,-44.3801380 &   18.29 & L1.5 &   54.5 $\pm$   0.2 &   736.7 &   238.6 \\%
WDS J15200-4423B  &                      & 230.0054769,-44.3798731 &   19.82 & L4.5 &   53.7 $\pm$   0.6 &   753.4 &   238.6 \\%
\hline
2MASS J16325610+3505076  & \multirow{2}{*}{57.1} & 248.2342852, 35.0851446 & L1.0 &  19.47 &    28.6 $\pm$   0.3 &   107.8 &   124.2 \\%
HD149361                 &                       & 248.2192979, 35.0750997 & K0V  &   8.03 &    29.0 $\pm$   0.0 &   107.4 &   125.6 \\%
\hline
2MASS J21265040-8140293  &            & 321.7115878,-81.6752636 & L3.0   &  20.72 &    29.2 $\pm$   0.9 &   128.5 &   153.9 \\%
TYC 9486-927-1           &    217.5   & 321.3662989,-81.6414894 & M1.0V  &  10.81 &    29.3 $\pm$   0.1 &   123.2 &   150.9 \\%
2MASS J21192028-8145446  &   1022.2   & 319.8360962,-81.7628668 & \ldots &  14.65 &    29.0 $\pm$   0.1 &   126.0 &   153.3 \\%
2MASS J21121598-8128452  &   2045.7   & 318.0681165,-81.4797055 & M5.5   &  14.04 &    28.6 $\pm$   0.1 &   123.8 &   155.0 \\%
\hline
DENIS-PJ220002.05-303832.9A & \multirow{2}{*}{1.0} & 330.0096692,-30.6428312 & M9.0 &  18.44 &    25.4 $\pm$   0.4 &   247.2 &   104.9 \\%
DENIS-PJ220002.05-303832.9B &                      & 330.0096946,-30.6425580 & L0.0 &  19.04 &    25.3 $\pm$   0.5 &   253.7 &   105.6 \\%
\hline
ULAS J230818.73+062951.4  & \multirow{2}{*}{3.8} & 347.0781929,6.4973599 & \ldots &  18.06 &    24.7 $\pm$   0.3 &   118.5 &   162.3 \\%
2665079816223169664       &                      & 347.0788922,6.4981654 & \ldots &  13.47 &    24.1 $\pm$   0.1 &   119.8 &   160.5 \\%
\hline
2MASS J23225299-6151275  & \multirow{2}{*}{16.6} & 350.7215915,-61.8579914 & L2.5 &  20.68 &    23.2 $\pm$   1.0 &   114.6 &   135.7 \\%
2MASS J23225240-6151114  &                       & 350.7191431,-61.8535236 & M5   &  14.90 &    23.6 $\pm$   0.1 &   110.3 &   135.2 \\%
\hline

\hline
\end{tabular}
\end{table*}

We searched for resolved binaries using the following criteria: 
\begin{equation} \label{eq1}
\begin{split}
\rho         & < 100 \varpi, \\
\Delta\varpi & < {\rm max}[1.0,3\sigma_{\varpi}],\\
\Delta\mu    & < 0.1\mu,\\
\Delta\theta & < 15\degr,
\end{split}
\end{equation}

\noindent where $\rho$ is the separation on the sky in arcseconds,
$\Delta\varpi$ is the difference of the \Gcat\ and candidate primary
parallaxes, $\varpi$ and $\sigma_{\varpi}$ are the parallax and error of the
\Gcat\ object, $\Delta\mu$ is the difference of the total proper motions, and
$\Delta\theta$ is the difference of the position angles. The chosen $\rho$
criterion is equivalent to 100,000\,au, which is a conservative upper limit
for a projected physical separation ($s$). This will meet the binding energy
criterion of $ |U_g^*| = G M_1 M_2 / s > 10^{33} J $ as developed by
\cite{2009A&A...507..251C} for a 0.1~M$_{\odot}$ + 2~M$_{\odot}$ system
\citep[see also][]{2010AJ....139.2566D}. The $\Delta\varpi$ criterion is based
on a consideration of the errors, standard 3$\sigma$ criterion or 1.0\,mas, to
allow for solutions that had unrealistically low errors. For the modulus and
position angles of the proper motion, criteria based on the errors would
remove nearby objects with significant orbital motion, hence we simply choose
hard criteria of $\sim$10\% in both parameters. This is large enough to
accommodate most orbital motion, but small enough to avoid false positives.
As discussed in Section~\ref{section:duplicatematches} two secondaries in
known wide binaries are missed by our criteria -- J0235-2331 (GJ 1048 B) and
J1442+6603 (G239-25B). We believe that in both cases the orbital motion
accounts for a $>10\%$ discrepancy in the proper motion criteria.

There are \NBINTOTFOUND\ objects in \NBINSYSTOT\ multiple systems including at
least one of our \Gcat\ objects. We compared this list to a combination of the
binary lists from the following publications:
\cite{2001AJ....122.3466M, 2014ApJ...792..119D, 2014MNRAS.445.3694D,
  2015AJ....150...57D, 2015ApJ...804...96G, 2015MNRAS.454.4476S,
  2016A&A...587A..51S, 2016ApJS..224...36K, 2017MNRAS.466.2983G,
  2017MNRAS.467.1126D}, and we found that \NBINPUBFOUND\ are known systems and
\NBINNEWFOUND\ are new systems. We found two systems, WDS~J15200-4423AB and
DENIS-P~J220002.05-303832.9AB, that were known spectroscopic binaries that
\G\ resolves.  In Table \ref{BinarySystems} we list systems that are
particularly worthy of discussion. Several of them include primaries with no
previous discussion in the literature, and are therefore identified with their
\G\ ID.

\begin{itemize}
\item  SDSS~J12451496+1204423 \citep{2010MNRAS.404.1817Z} is found to be a wide
companion ($s \geq 7900$\,au) to the DA white dwarf SDSS J124520.60+120531.3
\citep{2013ApJS..204....5K}. L dwarf + white dwarf non-interacting systems are
precious benchmarks, since the white dwarf can provide accurate age
constraints \citep[see e.g.][]{2011MNRAS.410..705D}.

\item  2MASS~J21265040-8140293 was identified by \cite{2016MNRAS.457.3191D} as a
companion to the young M dwarf TYC 9486-927-1. Analysis of the primary's
spectrum performed by \cite{2016MNRAS.457.3191D} revealed \ion{Li}~{\sc i}
$\lambda$6708\,{\AA} absorption consistent with an age range of 10--45\,Myr,
implying a mass range of 11.6--15\,$M_{\rm Jup}$ for the secondary. With a
projected separation of $\sim$7400\,au, 2MASS~J21265040-8140293 is the widest
orbit planetary-mass object known \citep{2018Geosc...8..362C}. Here we report two new candidate members of
this system, namely 2MASS~J21192028--8145446 and 2MASS~J21121598--8128452. Of them,
2MASS~J21121598--8128452 was classified as M5.5 \citep{2015ApJ...798...73G},
and would be the widest component of the system, with a projected separation
of $\sim$62700\,au. No spectral classification is given for
2MASS~J21192028--8145446, but since it is 0.61\,mag fainter than
2MASS~J21121598--8128452 we expect it to be an m6--7 dwarf (lower case
spectral type as this is a photometric estimate). Its projected
separation from the M1 primary is $\sim$31000\,au. 

We can compute a lower limit for the binding energy using the known spectral
types to estimate masses. For the M1 primary we assume a mass of 0.53
$M_\odot$, and for the M5.5 a mass of 0.1 $M_\odot$, by interpolating the
updated version of Table 5 from \citet{2013ApJS..208....9P}\footnote{\url{http://www.pas.rochester.edu/~emamajek/EEM_dwarf_UBVIJHK_colors_Teff.txt}}.
For the m6--7 dwarf, at the age of the system, the \cite{2003A&A...402..701B}
isochrones predict a mass in the 35--75 $M_{\rm Jup}$ range. We assume the
upper limit of this mass range in the following analysis. For the L3 dwarf we
adopt a mass of 15 $M_{\rm Jup}$, i.e. the upper limit of the range estimated
by \cite{2016MNRAS.457.3191D}. We also conservatively assume the semi-major
axis ($a$) to be equal to the observed projected separation (while in reality
$s \leq a$).  Under the above assumptions, the total binding energy for the
system would be $U_g^* \gtrsim -1.5 \times 10^{33} J$, so the system would
only be loosely bound \citep[see e.g.][Figure 1]{2009A&A...507..251C} and
unlikely to survive Galactic tides. We can determine an expected lifetime for
such a system using Equation 18 from \cite{2010AJ....139.2566D}. We find that
for the M5.5 the expected lifetime is $\sim$2.9\,Gyr, and for the m6--7 is
$\sim$5.8\,Gyr.

An alternative explanation would be that these are simply members of the same
young moving group. All four of these objects have indeed been selected as
candidate members of the Tucana--Horologium Association by
\cite{2015ApJ...798...73G}, while \cite{2016MNRAS.457.3191D} argue that
2MASS~J21265040-8140293 and TYC 9486-927-1 are members of the $\beta$~Pictoris
moving group. However, using \G\ updated astrometry and the BANYAN~$\Sigma$
online tool \citep{2018ApJ...856...23G}, we find 0\% Tucana--Horologium and
$\beta$~Pictoris membership probability. The main reason for this discrepancy
is probably that the four objects are approximately 5\,pc further away than
estimated using photometry in \cite{2015ApJ...798...73G} and
\cite{2016MNRAS.457.3191D}. Their \G\ proper motions on the other hand are
consistent
with the values used in those papers. Moreover, the initial membership
assessments were conducted using BANYAN~II \citep{2014ApJ...783..121G}, and
BANYAN~$\Sigma$ is known to provide more accurate membership probabilities
\citep{2018ApJ...856...23G}.

We find a non-zero probability membership only for the AB Doradus moving
group, with probability in the range 4.5--10.5\%, but the reported age range
for the system (10--45\,Myr) is inconsistent with the age of AB Doradus
\citep[100--125\,Myr;][]{2005ApJ...628L..69L}. We expect tools such as
BANYAN~$\Sigma$ to undergo a major overhaul following \DR\, with the
astrometry provided by \G\ strongly constraining the group kinematics. Further
discussion of the true nature of this association is therefore deferred to a
future paper.

\item  Four systems consist of members of young moving groups and
associations. 2MASS~J01415823--4633574 forms a wide common-proper-motion pair
with the M5.5 2MASS~J01443191--4604318. Both objects are members of the
Tucana--Horologium Association \citep[with 99.5\% and 99.8\% membership
  probability, respectively;
][]{2015ApJ...798...73G}. 2MASS~J02235464--5815067, 2MASS~J02251947--5837295,
and UCAC4~159-002053 are also members of the Tucana--Horologium Association
(with membership probability of 99.9\%, 99.7\%, and 99.9\%
respectively). 2MASSI~J0518461--275645 and 2954995674982867968 are both
members of Columba (99.9\% membership probability for both). These are very
wide systems, with typical projected physical separations, $s >$50,000\,au, and so
the nature of these systems is uncertain. Finally, 2MASS~J23225299--6151275
and 2MASS~J23225240--6151114 are also members of the Tucana--Horologium
Association (with membership probability of 96.7\% and 99.9\%, respectively),
but form a much tighter pair with projected physical separation of
$\sim$710~au. This system is therefore unequivocally bound.
\end{itemize}

These systems will provide valuable benchmark systems to constrain atmospheric
models and retrieval techniques. However, we have not tried to produce a
complete catalogue of binary systems containing UCD objects. As discussed in
Section~\ref{section:catalogue} our criteria fails for the binary systems GJ
1048 A/B and G 239-25 A/B in both cases because the modulus of the difference
in proper motions is greater than 10\%.  Hence the production of a complete
catalogue will require more sophisticated procedures, such as taking into
account the orbital motions of the components based on their predicted masses
and distances.


\section{Photometric Examination}
\label{section:PhotometricExamination}

\subsection{Absolute $G$ vs. $G-\grp$}
\begin{figure}
\begin{center}
\includegraphics[scale=0.5]{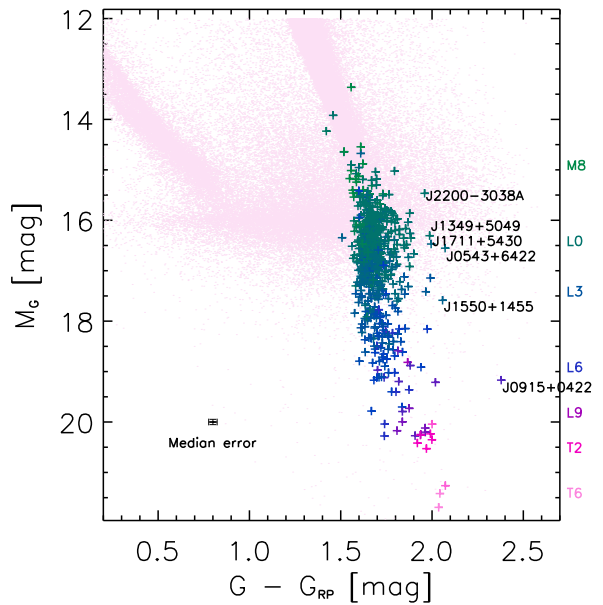}
\caption{Colour-magnitude diagrams for absolute $G$ vs. $G-\grp$. The light
  green points are all objects nominally within 100\,pc brighter than $M_G =
  12$\,mag from \DR\ regardless of quality flags, to
  delineate the white dwarf and main sequences. The crosses are the \Gcat\
  entries colour-coded by spectral type as indicated on the right hand side at
  the respective absolute magnitudes. Plotted in the lower left are median
  error bars.}
\label{Ggraphs}
\end{center}
\end{figure}
The most complete set of magnitudes for our UCD objects is in the
\G\ passbands, and these are also a new set of bands for studying these
objects. In Figure~\ref{Ggraphs} we plot the \DR\ absolute magnitude $M_G$
vs. colour $G-\grp$.

The $G-\grp$ colour shows a tight correlation that gradually increases from
1.5 to 2.1\,mag as one descends the main sequence. The standard deviation in
colour per absolute magnitude bin varies from 0.06 to 0.13\,mag. In the
\DR\ there are no published magnitude uncertainties to underline to the user
that the magnitude uncertainties are not symmetric. We have transformed the
flux uncertainties into magnitude upper and lower bounds and found a median
error of 0.02\,mag, indicating that the majority of the observed standard
deviation is due to intrinsic variations, which is in line with the intrinsic
spread seen in similar relations
\citep{2015ApJ...810..158F, 2016ApJS..225...10F}.

There are a number of outliers in Figure~\ref{Ggraphs}. In particular, there
are six UCD outliers that are 3$\sigma$ from the ``main-sequence''  
locus. We label them in the figure, and discuss them below: \begin{itemize}
\item J0543+6422 (2MASS J05431887+6422528) was spectroscopically found to be
  non-binary in \cite{2014ApJ...794..143B}. However, in the \DR\ there is an
  object detected (287767756635519488) at a separation of 0.6\arcsec, slightly
  brighter ($G$=18.96 vs. 18.97\,mag) and slightly redder ($G-\grp$ = 2.11
  vs. 2.07\,mag) but with no parallax estimate.  The uncertainty in position
  is very high (20.8 vs. 1.1\,mas in declination), consistent with a nearby
  object that is being constrained to having a zero parallax. The number of
  observations is however very different, 42 vs. 191, indicating that it may
  be the same object with observations assigned to two \dt{Source\_ID}s. The
  red colour and similar magnitude are consistent with both being an
  equal-mass binary with a separation of 0.6\arcsec\ or a single source with
  two \dt{Source\_ID}s.  There is no most probable scenario for this object
  and it is a prime candidate for observation with a ground-based adaptive
  optics system to confirm if it is actually a binary system.
\item J0915+0422 (2MASS 09153413+0422045) is a binary system of two L6 dwarfs
  with a separation of 0.73\arcsec\ \citep{2006AJ....132..891R}, at a distance
  of 18\,pc.  In the \DR\ data there is the probable match (\dt{Source\_ID}
  579379032257250176) 0.3\arcsec\ from the \Gcat\ position, 579379032258066432
  at a separation of 0.6\arcsec\. and 579379027962863104 at a separation of
  3.3\arcsec. Neither of the more distant matches have full solutions and the
  object at 3.3\arcsec\ is not red ($G-\grp$ = 1.4\,mag), while the closer
  detection has only a $G$ magnitude.  The number of along-scan observations
  are 111, 80 and 80 for the probable, close and more distant match
  respectively -- this difference in the number of observations for objects
  close on the sky is large but may not be indicating anything other than the
  downloading of \G\ observations are complicated. All objects may be real
  and, in some scan angles, \G\ may resolve them and in others may not. The
  position uncertainties are very different. For example in declination they
  are 0.9, 25.6 and 1.3\,mas respectively. The high uncertainty is consistent
  with a nearby object that is being constrained to having a zero parallax.

The most probable scenario is that 579379027962863104 (at 3.3\arcsec) is a
background star or galaxy and 579379032258066432 (at 0.6\arcsec) is the binary
companion from Reid at al. (2006), but it could also turn out that the
changing scanning direction correlates with the separation, and the matching
of the observations were compromised -- hence the lower number of
along-scan observations.  The source of the larger $G-\grp$ colour for this
system compared to a normal L6 is because the $\grp$ and $\gbp$ magnitudes are
found from integration of the $\grp$ and $\gbp$ fluxes in $3.5 \times
2.1\arcsec$ windows, and there is no provision for multiple sources in the
same window \citep{2018A&A...616A...4E}. Therefore, an excess in $\grp$ for
close binary systems is expected.  Indeed, in the $\grp-J$ or $\grp-z$ colours
J0915+0422 does not stand out, which is expected if the system is made of
similar objects and not resolved in both passbands that make up the colour. If
we assume the system is an equal mass binary the $\grp$ of an individual
component will be 0.75\,mag fainter, which is consistent with the 0.6\,mag
offset from the main sequence in Figure~\ref{Ggraphs}. We therefore conclude
that the \G\ $\grp$ for this object is the total system magnitude rather than
the individual component magnitude.

\item J1349+5049 (2MASS J13492525+5049544) has no literature indication of
  binarity and there are no other \DR\ detections nearby. The only
  \DR\ indication that may suggest a non-single solution is that it has the
  highest goodness-of-fit statistic for the along-scan observations of 84 (a
  ``good'' value would be 3), and the highest astrometric excess noise value
  for this sample.
\item J1550+1455 (2MASS J15500845+1455180) is a known L3.5 + L4 system with a
  separation of 0.9\arcsec\ \citep{2009AJ....138.1563B}.  In the \DR\ there is
  a detection of an object (1192782134013894144) at that separation from
  J1550+1455, but it has no parallax, $\grp$, or $\gbp$ magnitudes. The
  position uncertainties are not very high and both the probable match and the
  companion have over 200 observations, so the two of them are probably
  real. The very red colour of J1550+1455 could be due to the $\grp$ magnitude
  including flux form both components.
\item J1711+5430 (NLTT 44368B) was predicted to be a companion to NLTT 44368,
  an M3 at 90.2\arcsec\ based on proper motions
  \citep{2014ApJ...792..119D}. In Table~\ref{tab1711} we report the
  \DR\ parallaxes and proper motions. While the values are close, the
  differences in proper motions are significant and these two objects do not
  pass our binarity test developed in Section~\ref{section:BinarySystems}. The
  difference in proper motion may be due to binarity in J1711+5430. However,
  apart from its red $G-\grp$ colour for its $M_G$ magnitude, there is no
  published indication of unresolved binarity, there are no other
  \DR\ detections nearby, and the only \DR\ parameter that may be indicating
  multiplicity is the  duplicate flag, which is set to 1.
  
\begin{table}
 \caption{J1711+5430 and NLTT 44368 \DR\ parameters.}
\label{tab1711}
\begin{tabular}{lccc}   
\hline                                      
 Name &  $\varpi $ & $\mu_{\alpha} \cos{\delta}$    & $\mu_{\delta}$   \\
      &   mas                &  mas\,yr$^{-1}$  & \ mas\,yr$^{-1}$  \\
\hline                                      
J1711+5430 & 22.06 $\pm$ 0.60 & -48.71 $\pm$ 1.70 & 206.73 $\pm$ 1.904\\
NLTT 44368 & 21.14 $\pm$ 0.04 & -61.62 $\pm$ 0.12 & 211.31 $\pm$ 0.092\\
\hline                                      
\end{tabular}
\end{table}

\item J2200-3038A, as noted in Section \ref{section:BinarySystems}, is the brightest
component of the M9 + L0 system DENIS-P~J220002.05-303832.9AB with a
separation of 1.1\arcsec\ \citep{2006AJ....131.1007B}. The second component
does not have $\grp$ or $\gbp$ magnitudes, and the $\grp$ flux of the primary
component probably is the combination of both elements.
\end{itemize}


\subsection{Absolute $G$ vs. $\gbp-G$}
\label{EPPROB}
\begin{figure}
\begin{center}
\includegraphics[scale=0.5]{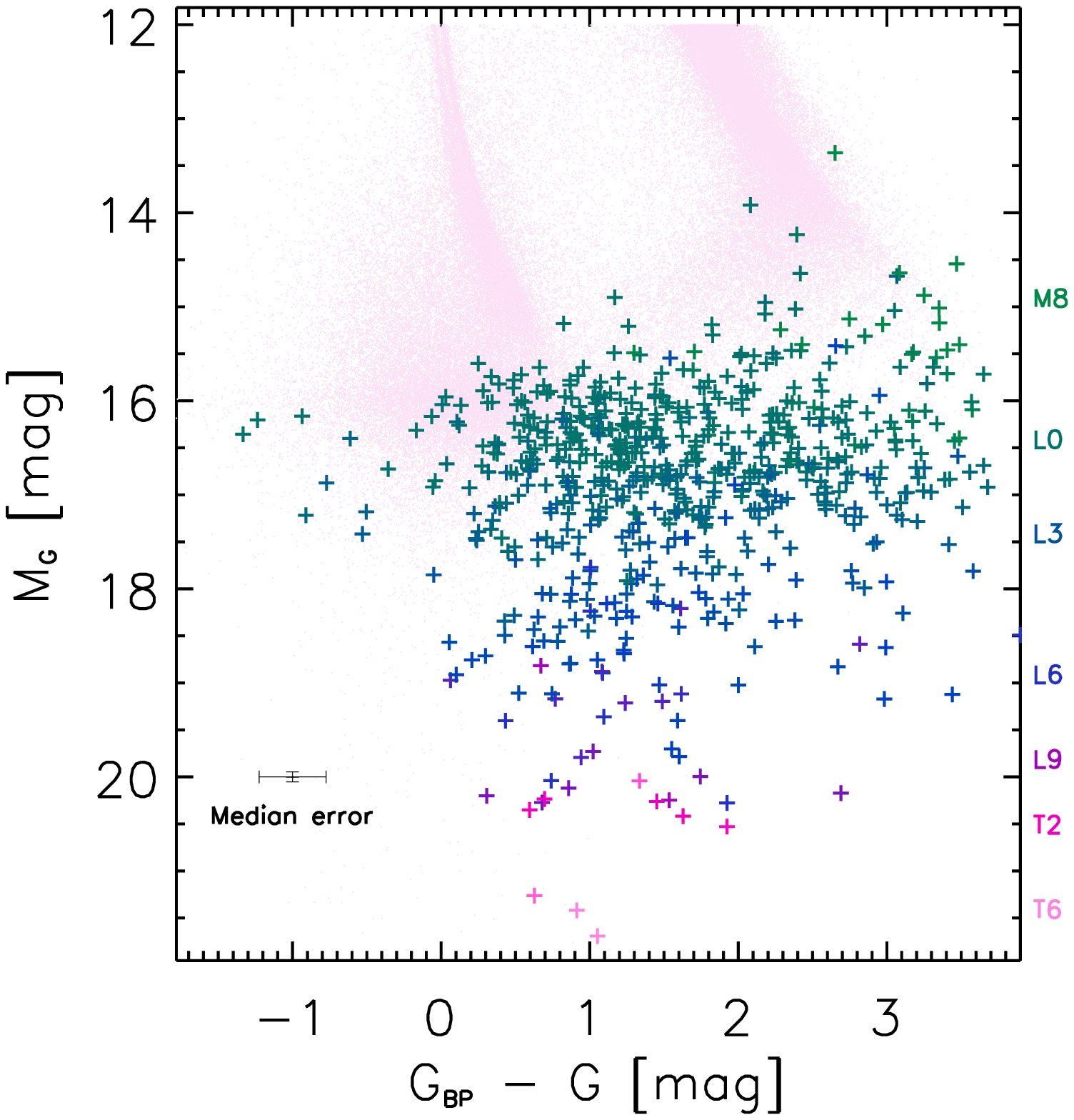}
\includegraphics[scale=0.5]{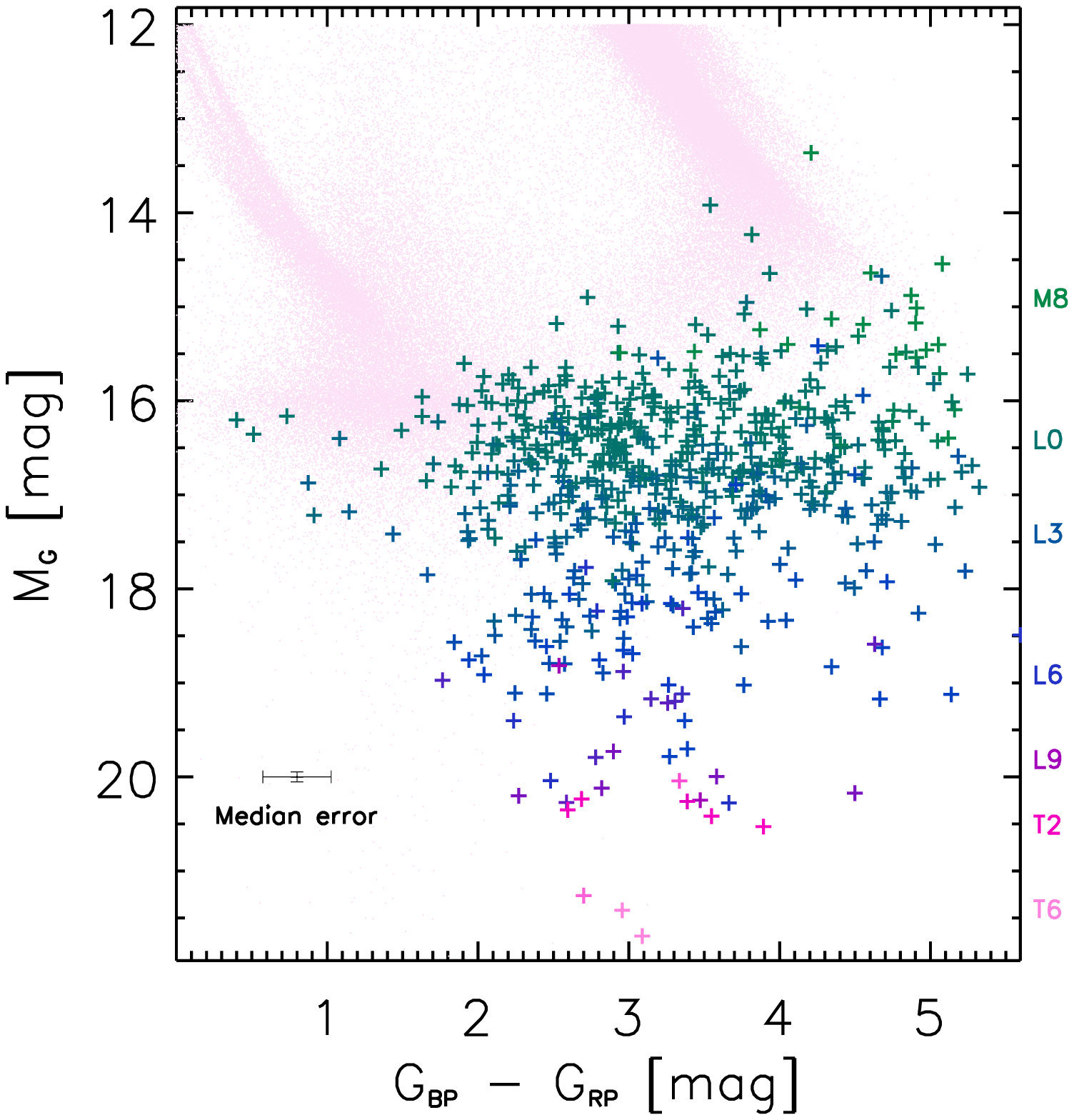}
\caption{Colour-magnitude diagrams for absolute $G$ vs. $\gbp-G$ ({\it top}) and
  $\gbp-\grp$ ({\it bottom}). The light pink points are all objects within
  100\,pc, the crosses are the \Gcat\ entries colour-coded by spectral type as
  indicated on the right hand side. Plotted in the lower left are error bars
  that are equivalent to the median uncertainty.}
\label{bpgraphs}
\end{center}
\end{figure}

\begin{figure}
\begin{center}
\includegraphics[scale=0.48]{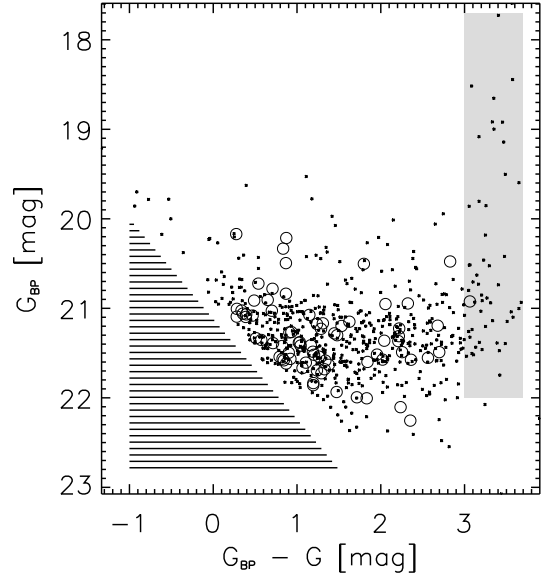}
\caption{{\it Top:} $\gbp$ vs. $\gbp-G$ for all objects. The hashed area is
  where objects are missed due to the $G$ band magnitude limit.  Open circles
  are objects with SDSS counterparts. The grey shaded area shows where we
  expect the $\gbp-G$ colour of these objects to occupy. }
\label{sdssbp}
\end{center}
\end{figure}

\begin{figure}
\begin{center}
\includegraphics[scale=0.48]{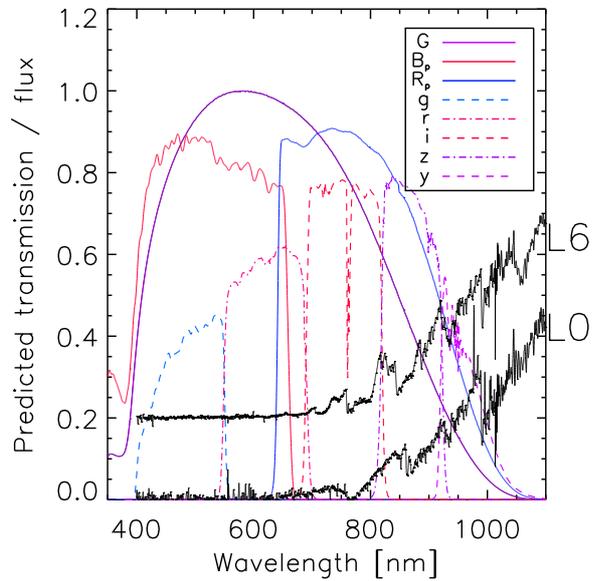}
\caption{Normalised filter and optical transmission for the \G\ and Gunn (used
  in the SDSS and PS1 surveys) passbands. Colours and passbands as indicated in
  the legend and normalisation of the two sets of filters are different and
  chosen to separate the two blocks.  The spectra are from
  X-Shooter for the L0 dwarf J2344-0733 ({\em bottom}) and the L6 dwarf J0006-6436 ({\em top}). }
\label{passbands}
\end{center}
\end{figure}

\begin{figure}
\begin{center}
\includegraphics[scale=0.48]{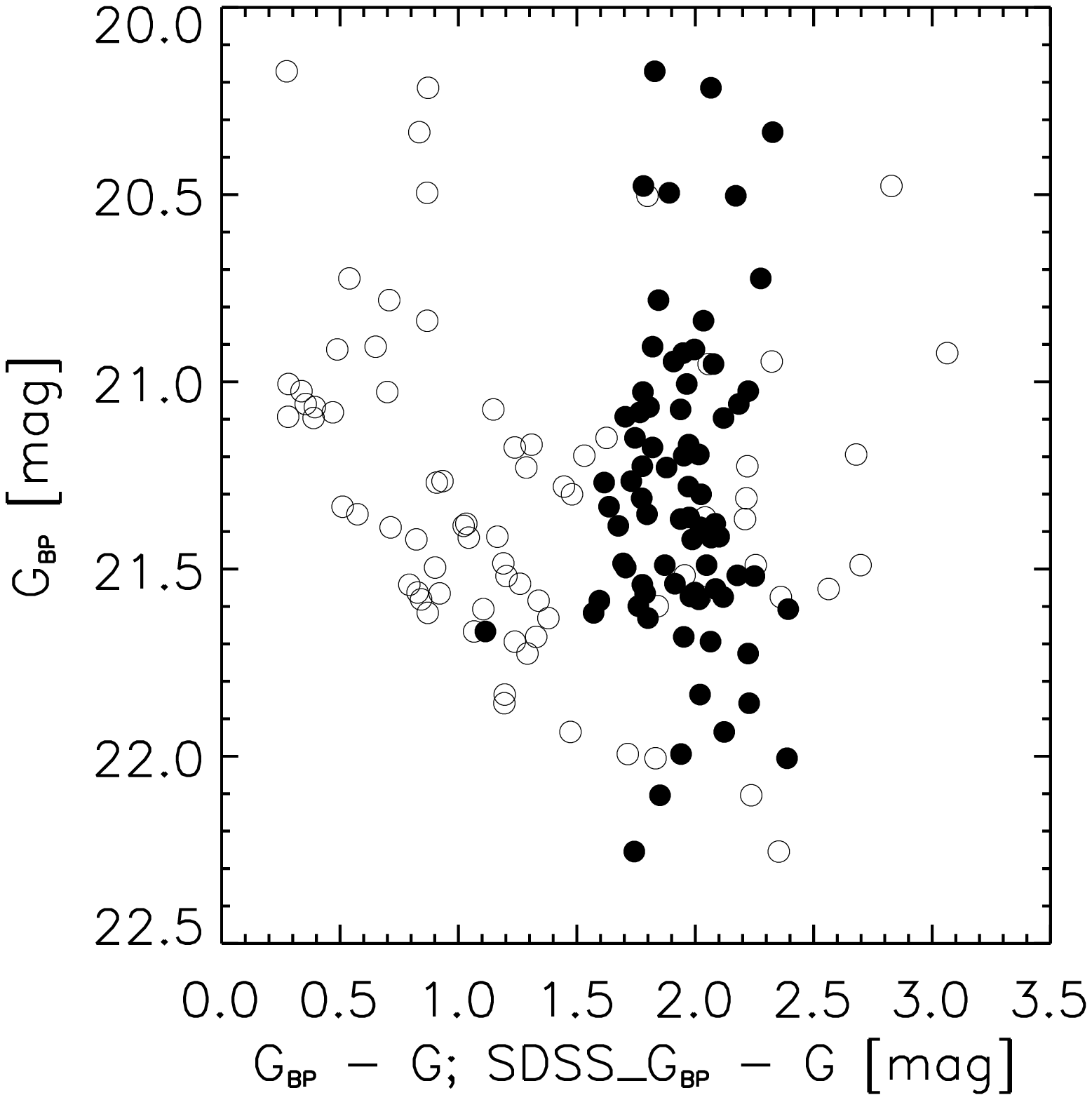}
\includegraphics[scale=0.48]{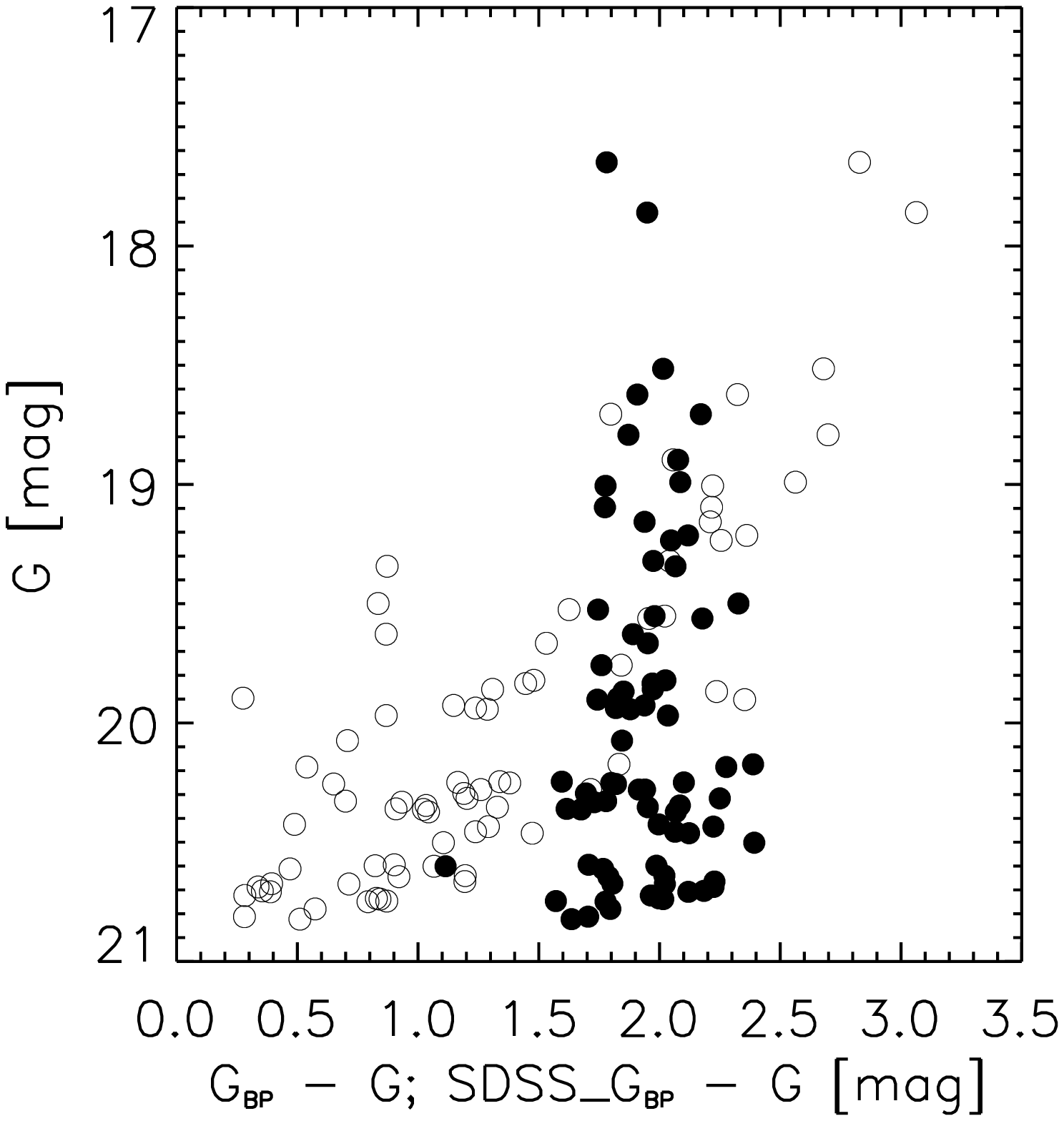}
\caption{{\it Top:} zoom on the objects with SDSS magnitudes from
  Figure~\ref{sdssbp}. Open circles as before using $\gbp$ vs. $\gbp-G$; filled
  circles are the same objects however plotting $\gbp$ vs. $SDSS\gbp-G$.  {\it Bottom:}
  same as top panel figure but plotting $G$ instead of $\gbp$ on the ordinate.
}

\label{sdssbpvsG}
\end{center}
\end{figure}

In Figure~\ref{bpgraphs} we plot $M_G$ vs $\gbp-\grp$ and $\gbp-G$ colours,
which have a strikingly higher dispersion relative to Figure~\ref{Ggraphs} for
a similar baseline in colour. The standard deviation in colour varies from 0.6
to 1.0 mag, while the median formal error is only 0.2\,mag. We cannot
assign this larger standard deviation to intrinsic variations as there is no
indications of this phenomenon in the literature for similar colour baselines.
In \citet[][]{2018A&A...616A..10G} they noted the larger scatter but merely
commented that these objects have very low flux in the $\gbp$ wavelength range,
making them intrinsically imprecise, which is evident in the comparison of the
three colour-magnitude plots. However, the scatter in Figure 4 is
present even for relatively bright UCDs, $\gbp \sim 19.5$,   and the uncertainties are not
consistent with such a large scatter.

Our sample is faint and, particularly in the blue band, at the limit of what
the \G\ team considers reliable photometry. If we apply the relative flux
error selection that \citet[][]{2018A&A...616A..10G} applied,
e.g. \dt{phot\_g\_mean\_flux\_over\_error} $>$50,
\dt{phot\_rp\_mean\_flux\_over\_error} $>$20, and
\dt{phot\_bp\_mean\_flux\_over\_error} $>$20, then of the 695, 660 and 660
objects with published magnitudes in the $G$, $\gbp$ and $\grp$ bands only
693, 14 and 602 would remain. In addition they constrained the flux ratio
$(I_\gbp + I_\grp )/I_G$ (\dt{phot\_bp\_rp\_excess\_factor}) to the range
$1.0+0.015 (\gbp-\grp )^2 <$ \dt{phot\_bp\_rp\_excess\_factor} $< 1.3 + 0.06
(\gbp-\grp)^2$, which would reduce our 660 sample to only 218.  Indeed for the
Figure 9 of \citet[][]{2018A&A...616A..10G} they did not apply this filter on
fluxes as the size of the sample would have been significantly reduced.

In \citet{2018A&A...616A..17A} they estimated a unit-weight
uncertainty\footnote{The ``unit-weight uncertainty'' is the ratio of the
  calculated unit weight and an independent estimate of the true error.}  of
1.3 assuming that the widths of main sequences in Galactic clusters were due solely
to photometric uncertainties. The large standard deviation of the $\gbp-G$
colour with respect to the median uncertainty implies a unit-weight
uncertainty of $\sim$3. Therefore, either there is a large intrinsic scatter
or the uncertainties of the $\gbp$ are significantly underestimated.

In Figure~\ref{sdssbp} we show the $\gbp-G$ colour versus the $\gbp$
magnitude for all UCDs. We expect the colour to be clustered at a $\gbp-G
\sim 3$\,mag, as outlined by the grey box. The brightest examples fall within
this range, but for $\gbp > 19.5$\,mag the UCDs appear to be spread evenly.
To investigate the possibility that the observed scatter is intrinsic we
examine the SDSS magnitudes. In Figure~\ref{passbands} we show that the $\gbp$
band coverage is roughly equal to the combined SDSS $g$ and $r$ coverage.  We
have taken those objects from our sample that have $g$ and $r$ magnitudes in
the SDSS, and constructed a pseudo-$\gbp$ magnitude, dubbed $SDSS\gbp$, by
adding the fluxes in the $g$ and $r$ SDSS bands.  We restricted the selection
to objects with uncertainties in $\gbp$, $r$ and $g$ to less than 0.6\,mag,
which provided a sample of 75 M9-L1 objects with $\gbp$ between 20.17\,mag and
22.25\,mag. The objects with SDSS counterparts are plotted as open circles in
Figure~\ref{sdssbp}.

In the top panel of Figure~\ref{sdssbpvsG}, the objects with $SDSS\gbp-G$
colours (filled circles) centre on $\sim$2\,mag with a dispersion of 0.2\,mag
that increases slightly as the objects get fainter. The $\gbp-G$ colours of
the same objects (open circles) show a lack of clustering with a dispersion of
0.70\,mag, even though the median error is 0.25\,mag.  The offset between the
$SDSS\gbp-G$ colour at $\sim$2\,mag and the predicted $\gbp-G$ colour at $\sim
3$\,mag is not unexpected, as the $g$ and $r$ passbands cover the same
spectral range as $\gbp$, but the combined profile is different. Besides, the
SDSS magnitudes are on the AB magnitude system, while the zero point of the
\G\ magnitudes are set by Vega.

Another indication of problems in the $\gbp$ passband for faint red objects
can be seen in Figure 33 of \citet{2018A&A...616A..17A}, where the main
sequence of the Alessi 10 cluster deviates from the expected path at $\gbp
\sim 19.5$\,mag.  As this cluster is considered a dense field they cited a
number of possible contributing factors (underestimated sky background,
overlapping spectra, extended objects and blended objects), but these factors
would not be appropriate for our targets, which are primarily in low density
regions.

\begin{figure}
\begin{center}
\includegraphics[scale=0.48]{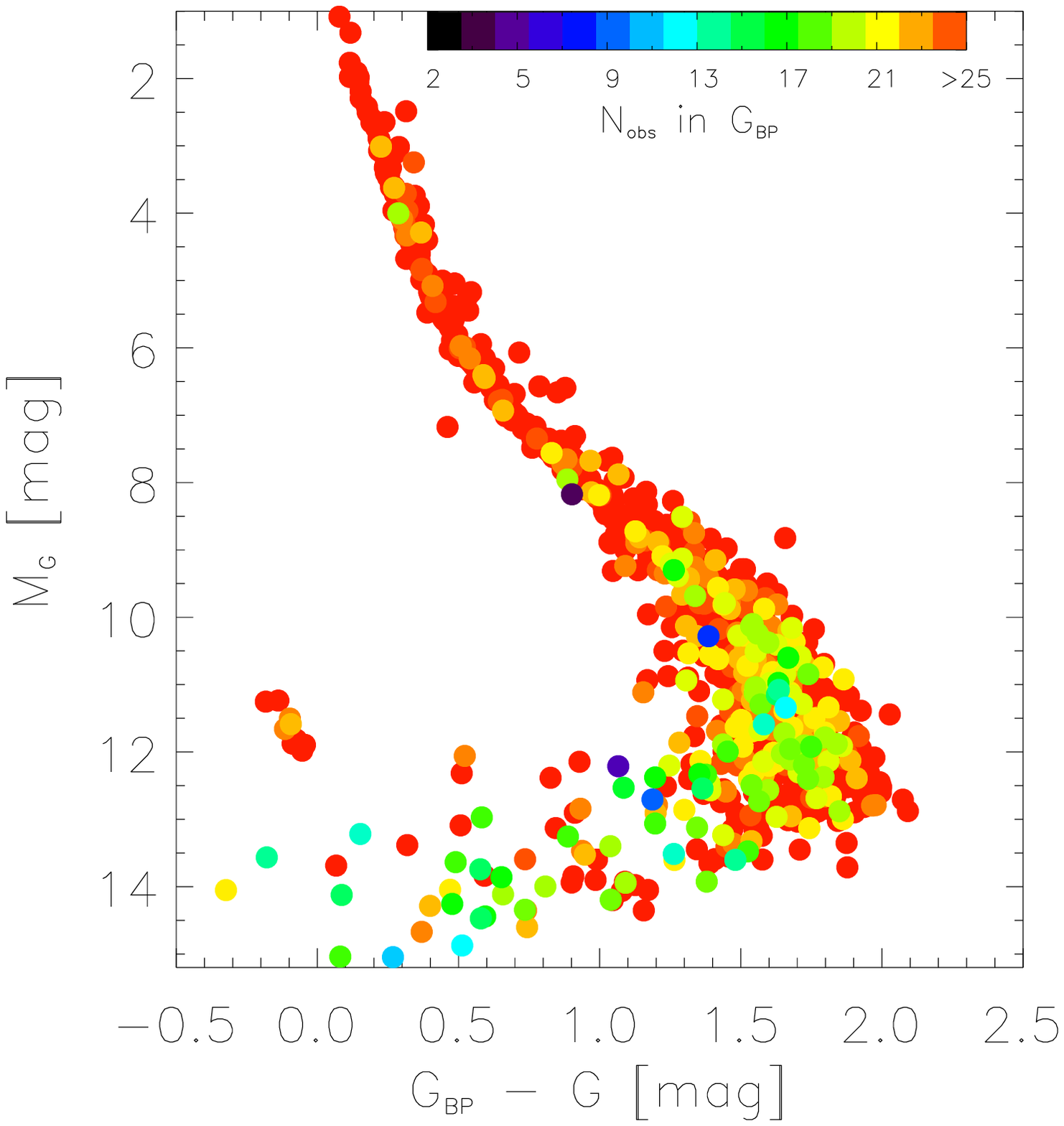}
\includegraphics[scale=0.48]{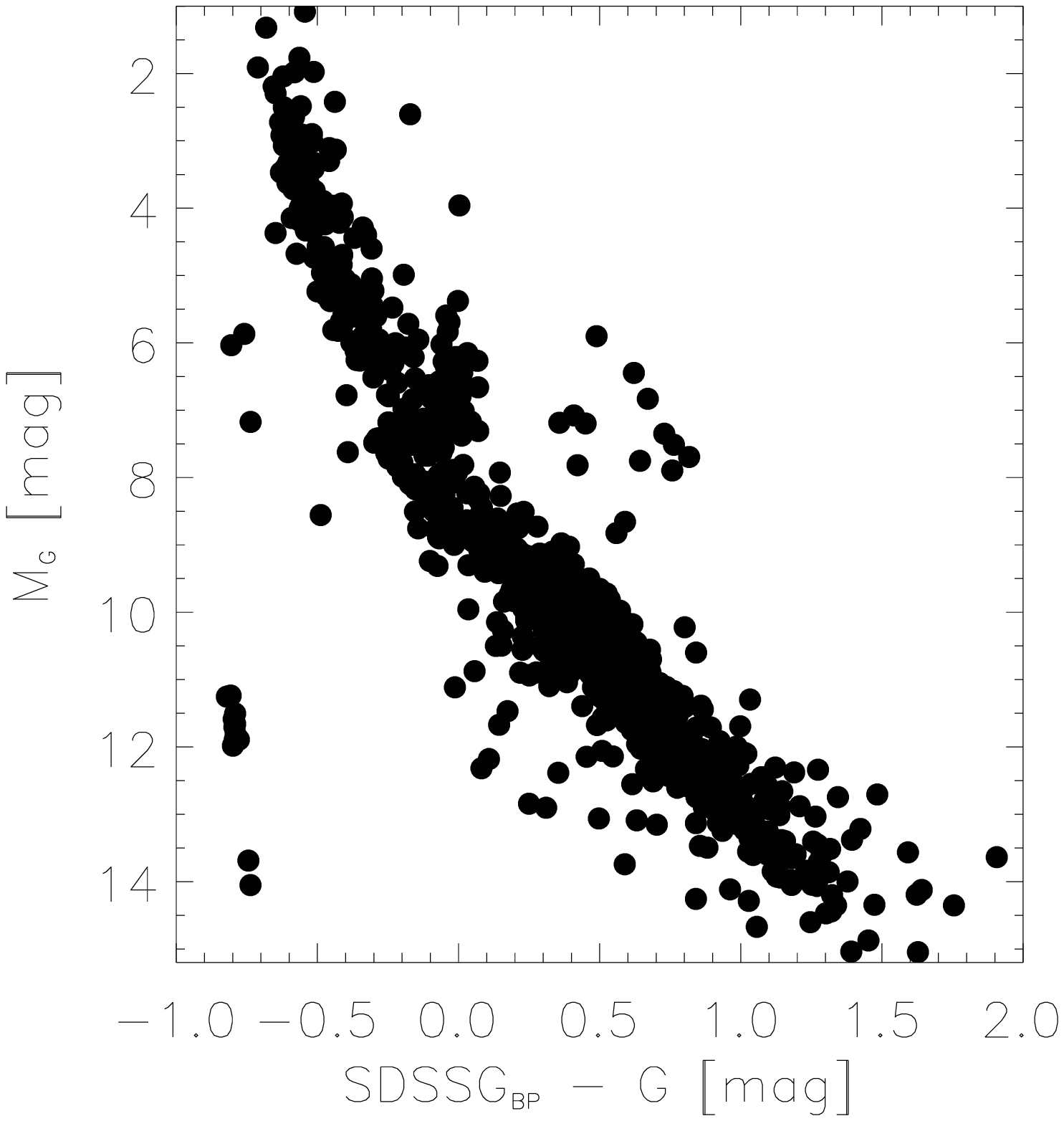}

\caption{{\it Top:} $M_G$ vs. $\gbp-G$ for members of Praesepe cluster
  selected astrometrically. The colour coding follows the number of
  observations in the $\gbp$ band as indicated in the colour bar.  {\it
    Bottom:} $M_G$ vs. $SDSS\gbp-G$ for the same sample of objects as the top panel.}

\label{Praesepe}
\end{center}
\end{figure}

In order to test the reliability of $\gbp$ in another cluster, we constructed
a sample of the Praesepe cluster members using only the astrometric parameters
in the \DR. We selected all objects with ($\alpha,\delta$) in the range
(126--135, 16--24)\degr, $\varpi $ in the range 3.--7. mas and ($\mu_{\alpha}
\cos{\delta}$, $\mu_{\delta}$) in the ranges (-30.---40., -10.---18.)\masyr\ 
based on the membership sample provided in \cite{2018A&A...616A..10G},
resulting in 1336 members listed here.  There was no limit made on the
quality of the photometry, as this would have removed all of the faint
members. This cluster was chosen as it has a proper motion that is
significantly different from the field so we can be quite confident that the
sample is dominated by Praesepe members.  In Figure~\ref{Praesepe} we plot
$M_G$ vs. $\gbp-G$ in the top panel, where a deviation of the main sequence
from the expected path for faint red objects is seen, as in
\citet{2018A&A...616A..17A} for Alessi 10.  The authors colour-coded the
Alessi 10 members by the number of observations in the $\gbp$ band, and
noted that the objects with the lowest number of observations are
predominantly in the deviated region. We have made the same colour coding in
Figure~\ref{Praesepe}, but the objects with lower numbers of observations are
not confined to the deviated part. More examples are required to see if the
the correlation of deviation with number of observations observed in Alessi 10
is significant.

In the lower panel of Figure~\ref{Praesepe} we plot the same objects using
$SDSS\gbp$ instead of $\gbp$. The spread in the main sequence is larger than
the top panel because the SDSS magnitudes are less precise; this is also a
very dense region that adversely impacts the SDSS measurements compared to
the \DR\ ones. The distinct discontinuity in the main sequence at
$M_G \sim$7.0\,mag is due to the brightest objects being saturated in the
SDSS. However, the lower main sequence in $SDSS\gbp$ follows an expected
increasingly redder path for  fainter objects not unexpected deviated path
of the top panel.

We examined other samples of selected red sources and found the $G-\gbp$ colour
was significantly noisier than the $SDSS\gbp-\gbp$ for the late type M dwarfs
catalog from \cite{2010AJ....139.1808S} but the colours are consistent for
early M dwarfs \citep{2011AJ....141...97W}, carbon stars
\citep{2004AJ....127.2838D}, white dwarfs \citep{2019MNRAS.482.4570G}, and
quasars \citep{2015ApJS..221...12S}. As a result, we find the $\gbp$ and
uncertainty values are inconsistent only for very red, faint, objects.

The $\gbp$ flux, from which the magnitude is derived, is the mean of the
integrated spectra in the aforementioned $3.5 \times 2.1\arcsec$ windows over
all the observations.  These objects are extremely faint in $\gbp$, many are
background-limited, and one possible reason for underestimating the $\gbp$
may be because the error of the mean is dominated by the variation of the
background flux, not by the variation of the objects flux. Another possibility
is the position of the geometric windows are placed for the $\gbp$ and $\grp$
filters using the \G\ $G$ position, and perhaps the very red colour leads to a
systematic offset in the $\gbp$ window position.

Since the $\gbp$ value comes essentially from aperture photometry, any
detection level is crucially dependent on the background determination. A
typical example of the differing fluxes can be seen in Table~\ref{allcat} for
J1807+5015. It has fluxes of 1420.5, 58.5 and 2676.0 erg cm$^{-2}$ s$^{-1}$
Hz$^{-1}$ for the $I_G, I_\gbp$ and $I_\grp$, respectively.  As our objects
are significantly above background in both the $G$ and $\grp$ passbands, the
simplest explanation is that a $\gbp$ magnitude is included when robust $G$
and $\grp$ detections are made, even if the $\gbp$ detection is not itself
significant, hence the derived $\gbp$ magnitudes are determined by the
background more than by the object. There is considerable complexity in the
derivation and calibration of \G\ magnitudes and we conclude that any use of
the $\gbp$ passband for faint red objects must be made with caution and do not
use it further for this work.

\subsection{Colour-magnitude diagrams using external magnitudes}
\label{PS1}

\begin{figure}
\begin{center}
\includegraphics[scale=0.42]{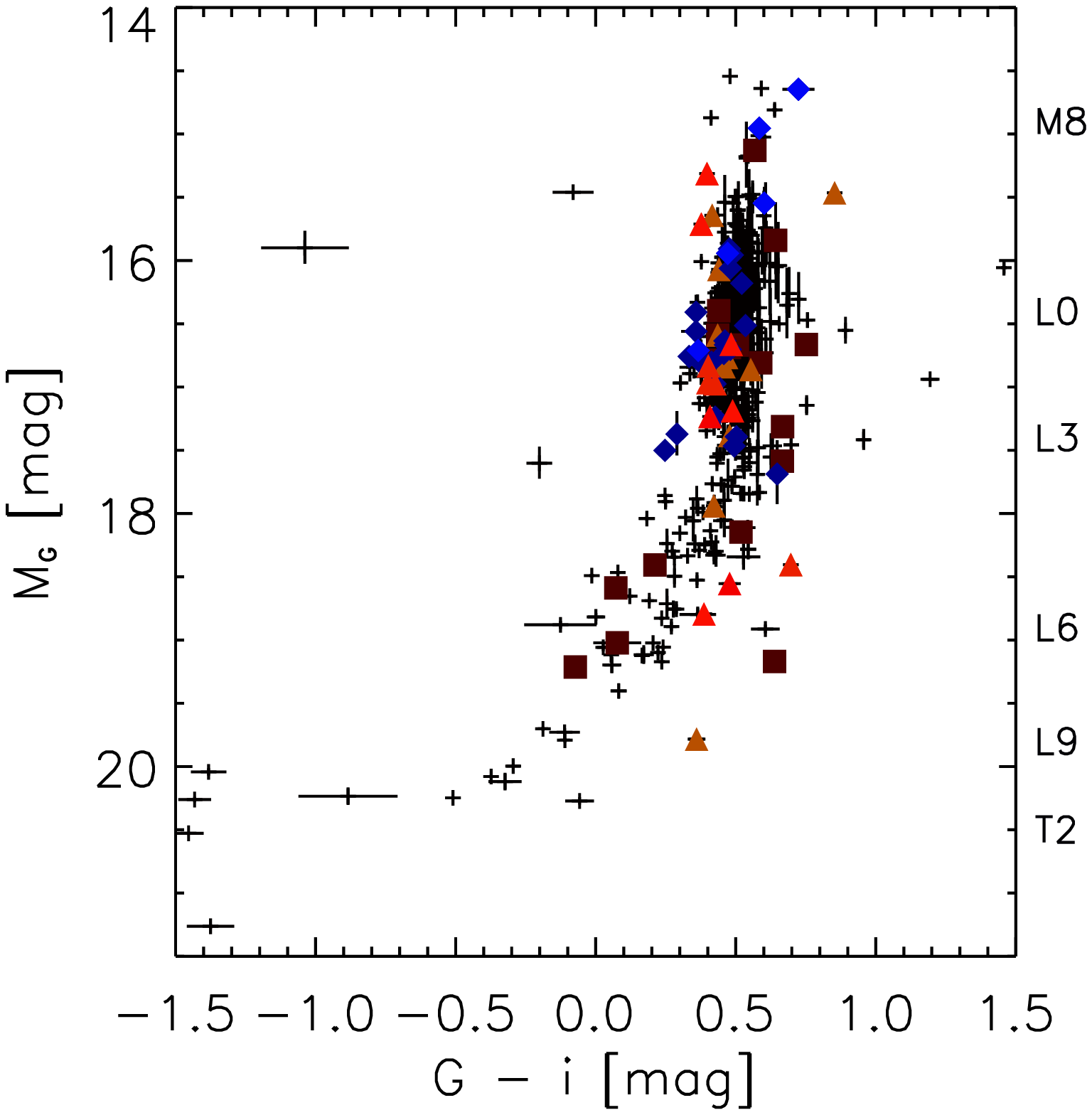}
\includegraphics[scale=0.42]{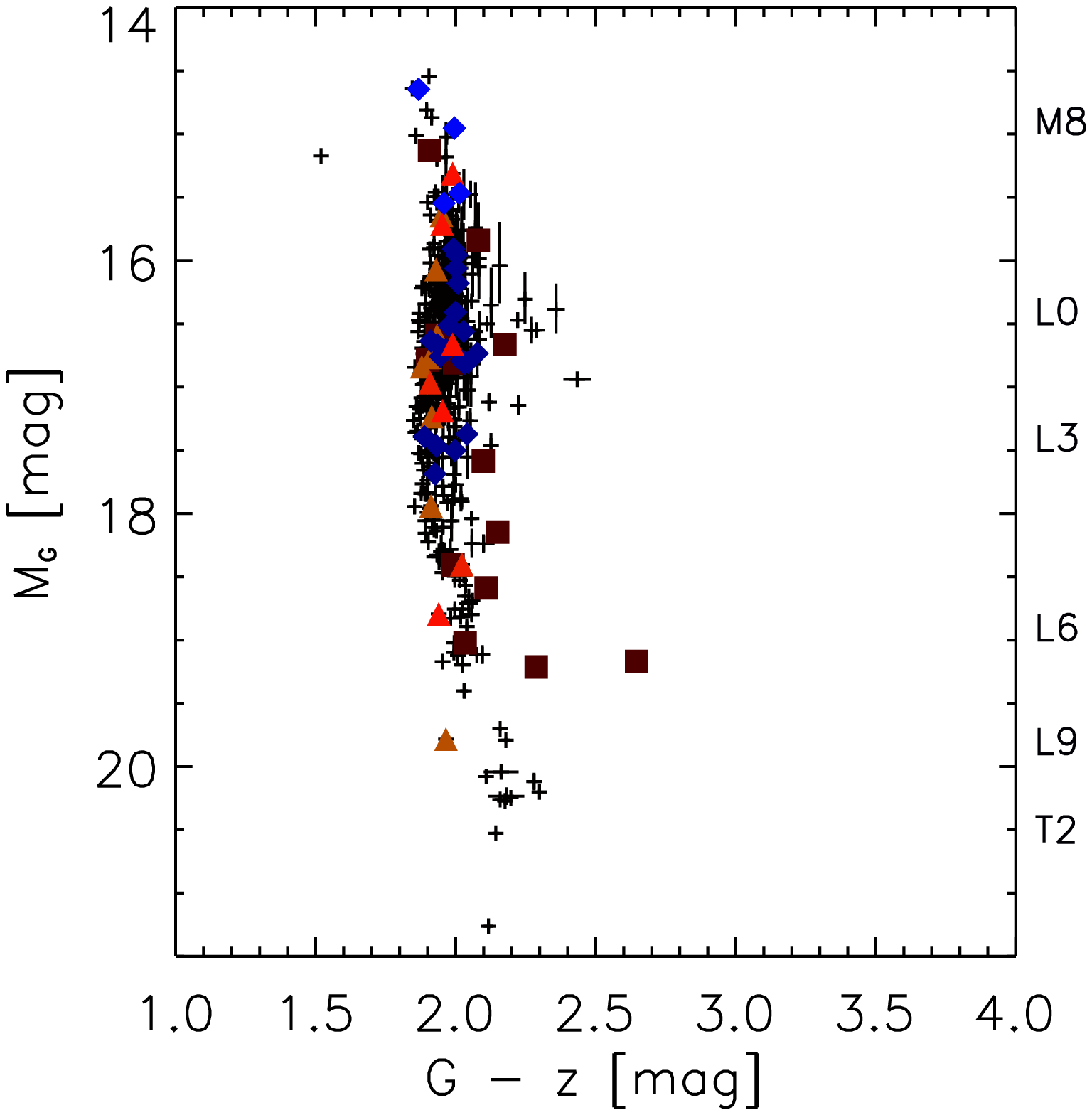}
\includegraphics[scale=0.42]{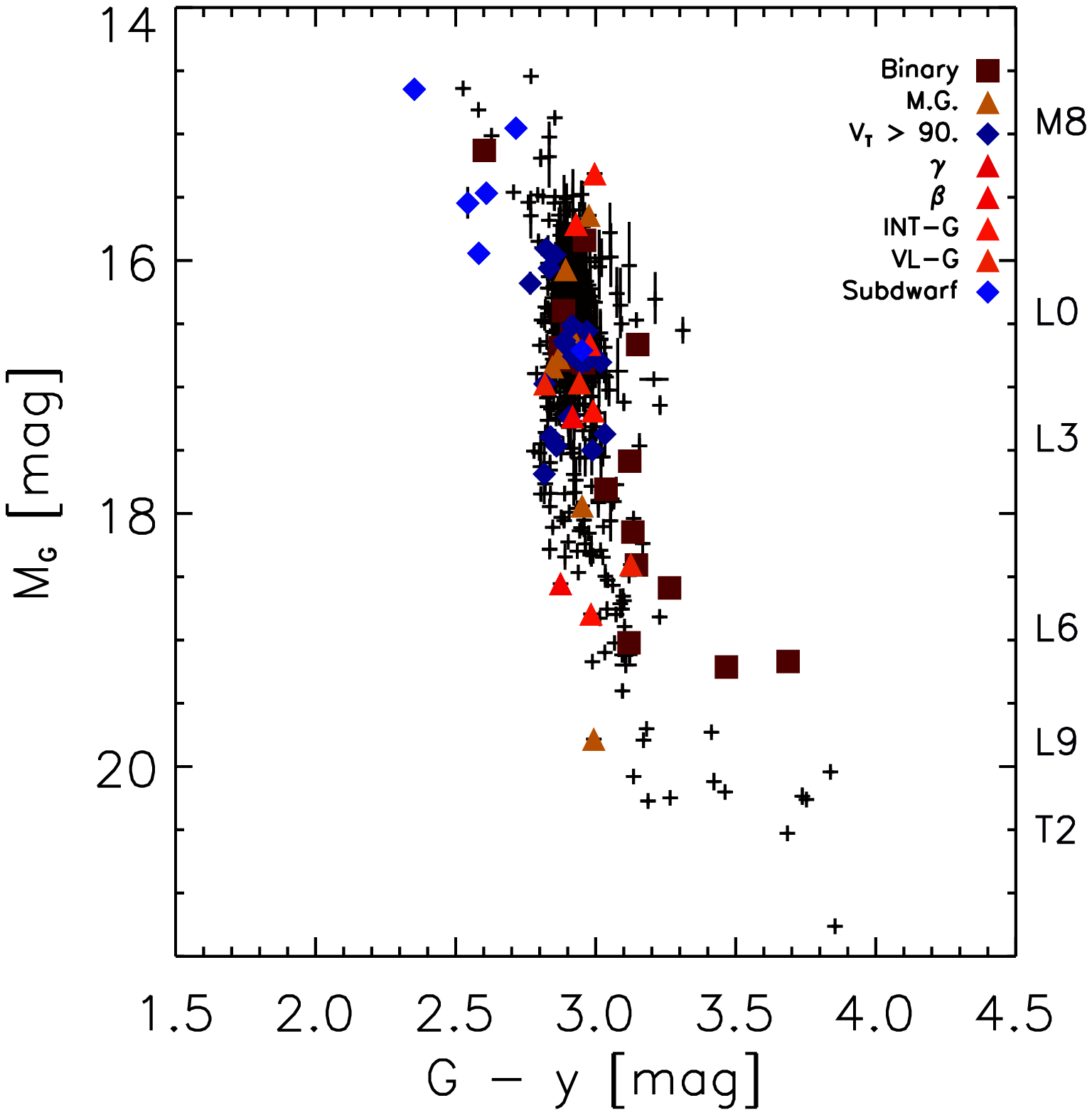}
\caption{Colour-magnitude diagrams. $G$-band absolute magnitude as a function
  of $G-i$, $G-z$ and $G-y$ from the Pan-STARRS PS1. The legend in the
  third panel indicates the symbols used for spectral type qualifications,
  binarity and high tangential velocity objects.}
\label{GUNNCMD}
\end{center}
\end{figure}

In Figures~\ref{GUNNCMD} through \ref{WISECMD} we plot the colour combinations
of the $G$ band and the PS1, 2MASS and AllWISE magnitudes versus absolute $G$
magnitudes for the \Gcat\ objects. Within each sequence of $M_G$ absolute
magnitude comparisons with external photometry we have set the relative range
on the axes to be the same to simplify inter-comparisons. In each graph we
have indicated on the left-hand axis the average spectral type corresponding
to the $M_G$ for the main bulk of stars. Old, young or binary systems do not
correspond to this scale. If we replace $G$ with $\grp$ the
overall trends do not change.

In the last panel of each sequence we indicate spectral typing qualifications
in the literature with the use of different symbols. For each entry in a
binary system we plot as brown squares those unresolved binaries or systems
with angular separations $\rho \le$1\arcsec\ on the assumption that ground-based
programs are unable to extract the magnitudes of the different components if
the separation is smaller.  This is not always the case: e.g. an \Gcat\ system
has a nominal separation of $\rho >$1\arcsec\ but it is not resolved and the
magnitude is a combined value: or, the \Gcat\ system has a $\rho <$1\arcsec\ but
the published magnitudes are of the separate components.  We assumed that entries
that have the gravity indicators $\gamma$, $\beta$ 
\citep{2009AJ....137.3345C}, ``int-g'' or ``vl-g''
\citep{2013ApJ...772...79A}, or that are confirmed members of known moving
groups are young and we have plotted them as upright triangles with colours as
indicated in the legend. Finally, we assumed objects listed as subdwarfs or
with $V_{\rm tan} > 90$\,km/s are old, and have plotted them as diamonds.

\subsubsection{\G\ and Pan-STARRS PS1 magnitudes}
We limit our examination for the PS1 catalogue to the $i, z$ and $y$
passbands, because we find that 50\% and 30\% of the values in $g$ and $r$,
respectively, have bad quality flags or do not have error estimates.  As shown
in Figure~\ref{passbands}, the $G$ band has significant sensitivity in these
three PS1 bands and from that plot the effective wavelength of $G$ appears to
be bluer than $r$. However, the effective wavelength is object-dependent and,
on average, for L dwarfs the $G$ band effective wavelength is very close to
that of $i$.

In Figure~\ref{GUNNCMD} the sequence for 15.5\,mag $< M_G <$ 18\,mag, roughly
L0 to L4, has remarkably constant $G-i$, $G-z$ and $G-y$ colours with widths
of 0.07--0.08\,mag.  The earlier M dwarfs and later LT dwarfs deviate to bluer
and redder colours, respectively. The objects with old and young spectral
characteristics have dispersions of 0.08\,mag in the $G-i$ and $G- y$ colours
and up to 0.04\,mag in $G-z$. Even though the overlap of all objects is quite
significant, there is some correlation with the old and young dwarfs, being
consistently on one side or the other of the main bulk of objects. For types
later than L6 ($M_G>$18\,mag) in the bluer $G-i$ and redder $G-y$ colours, the
deviations from the fixed colours of the earlier types reaches 0.5\,mag and
the trend increases with cooler spectral types. These colours can be useful
for spectral type differentiation of late L and T dwarfs.  Alternatively $G-z$
offers an almost constant value from $M_G$= 15 to 20\,mag.

\subsubsection{\G\ and 2MASS PSC magnitudes}

\begin{figure}
\begin{center}
\includegraphics[scale=0.43]{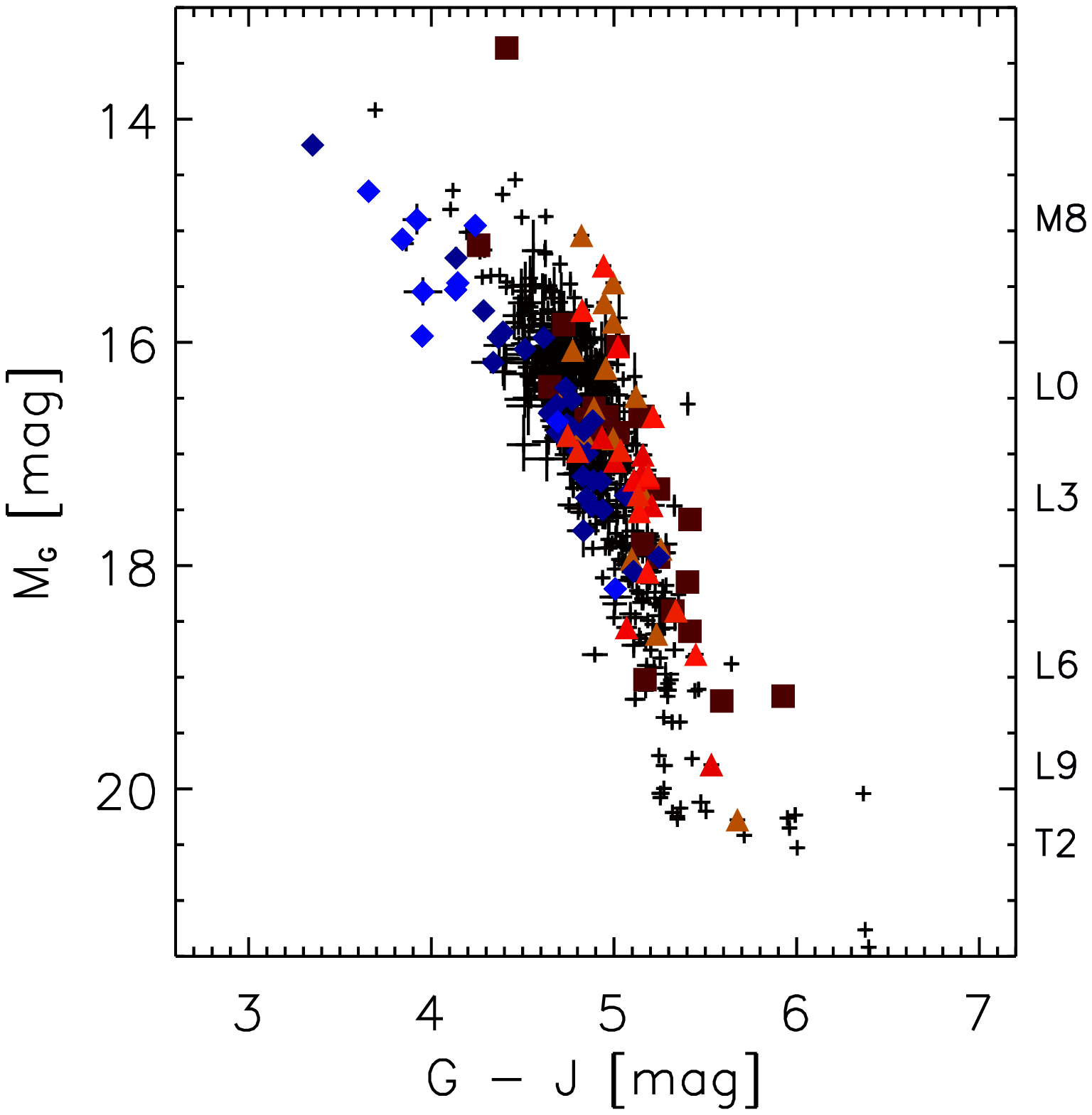}
\includegraphics[scale=0.43]{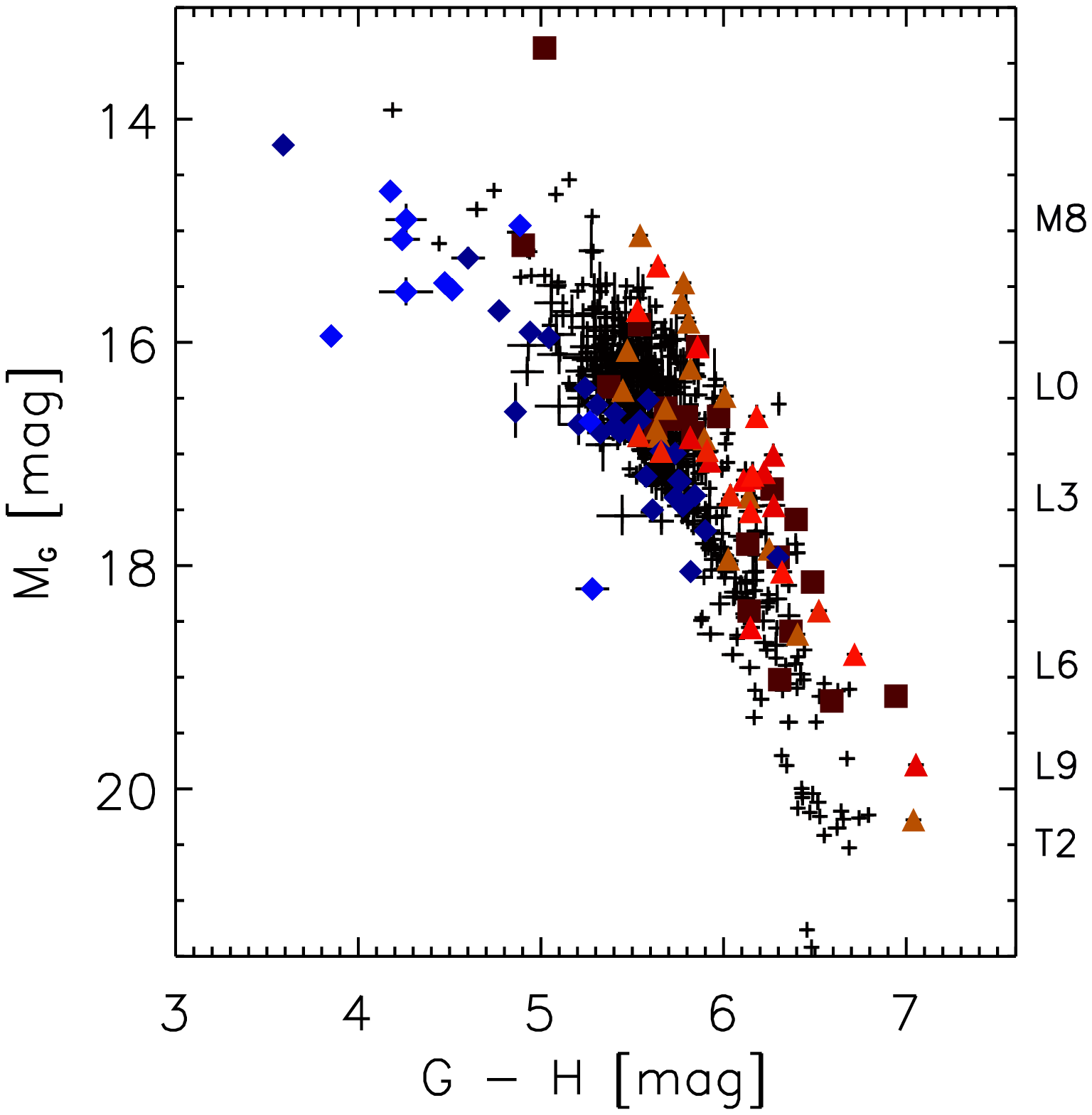}
\includegraphics[scale=0.43]{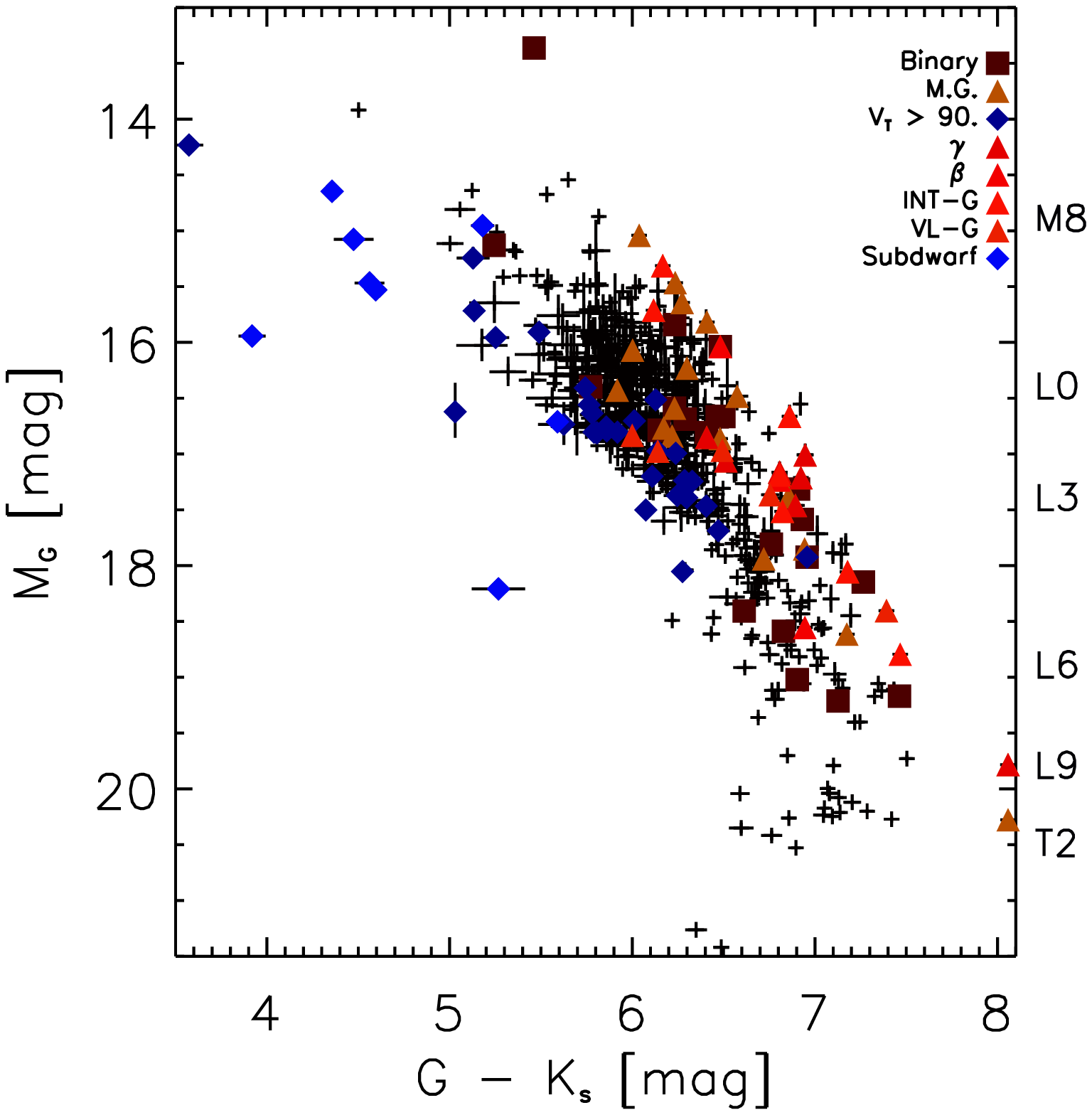}
\caption{Same as Figure~\ref{GUNNCMD} but for $G-J$, $G-H$ and $G-K_s$  colours from the
  2MASS PSC. }
\label{TMASSCMD}
\end{center}
\end{figure}

Figure~\ref{TMASSCMD} is the sequence of $G$ absolute magnitude comparisons
with $G$--2MASS colours. The mean colours vary by $\sim$1.5\,mag in all three
relations. The dispersion increases from 0.16, 0.23 and 0.30\,mag for the
$G-J$, $G-H$ and $G-K_s$, respectively. The mean colour for the old and young
samples separates by 0.6, 1.0 and 1.2\,mag for the $G-J$, $G-H$ and $G-K_s$
colours, respectively.  The underlying sequences maintain relatively linear
relations with increasing slopes as the baseline colours increases.  Overall,
for L dwarfs all three colours continue to get redder as the objects get
fainter in $G$. At the L-T boundary the three colours vary differently: redder
in $G-J$, unchanging in $G-H$, and a turn around to bluer colours in
$G-K_s$. The two ``young'' objects (J0355+1133 \citep{2009AJ....137.3345C} and
J2148+4003 \citep{2010ApJS..190..100K}) along with the bulk of other objects
with young indicators continue to move redward in all three colours.  A
primary cause of the increased spread in colours from $G-J$, through $G-H$, to $G-K_s$
plausibly corresponds to $H$- and $K_s$-band suppression from atomic and
molecular absorption of methane and H$_2$ collision-induced absorption
\citep[e.g.][]{2011MNRAS.414..575M}, which leads to relatively brighter $H$
and $K_s$ bands for the lower gravity young objects and in turn redder colours
relative to the higher gravity older objects.

%

\subsubsection{\G\ and AllWISE magnitudes}

\begin{figure}
\begin{center}
\includegraphics[scale=0.43]{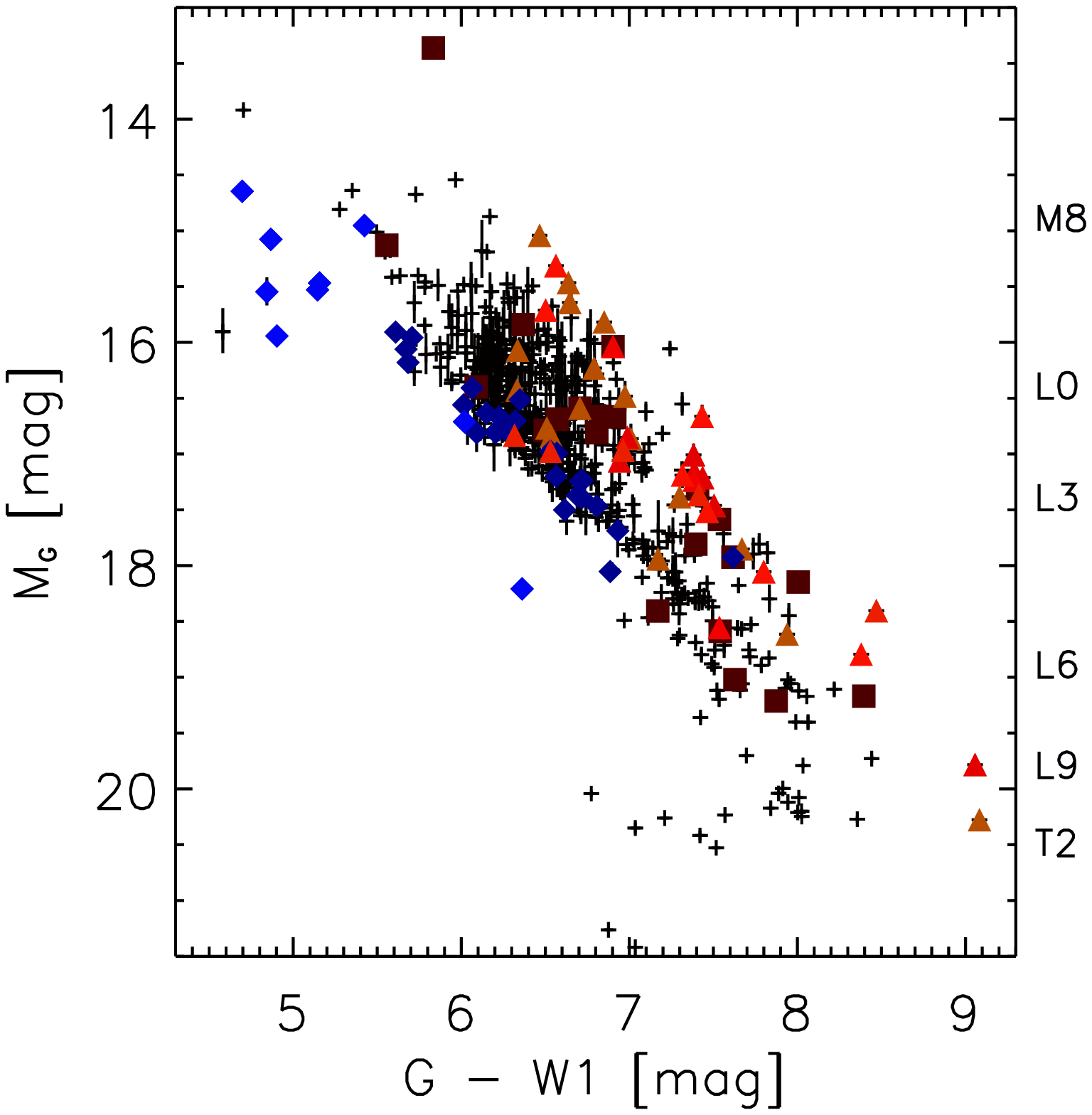}
\includegraphics[scale=0.43]{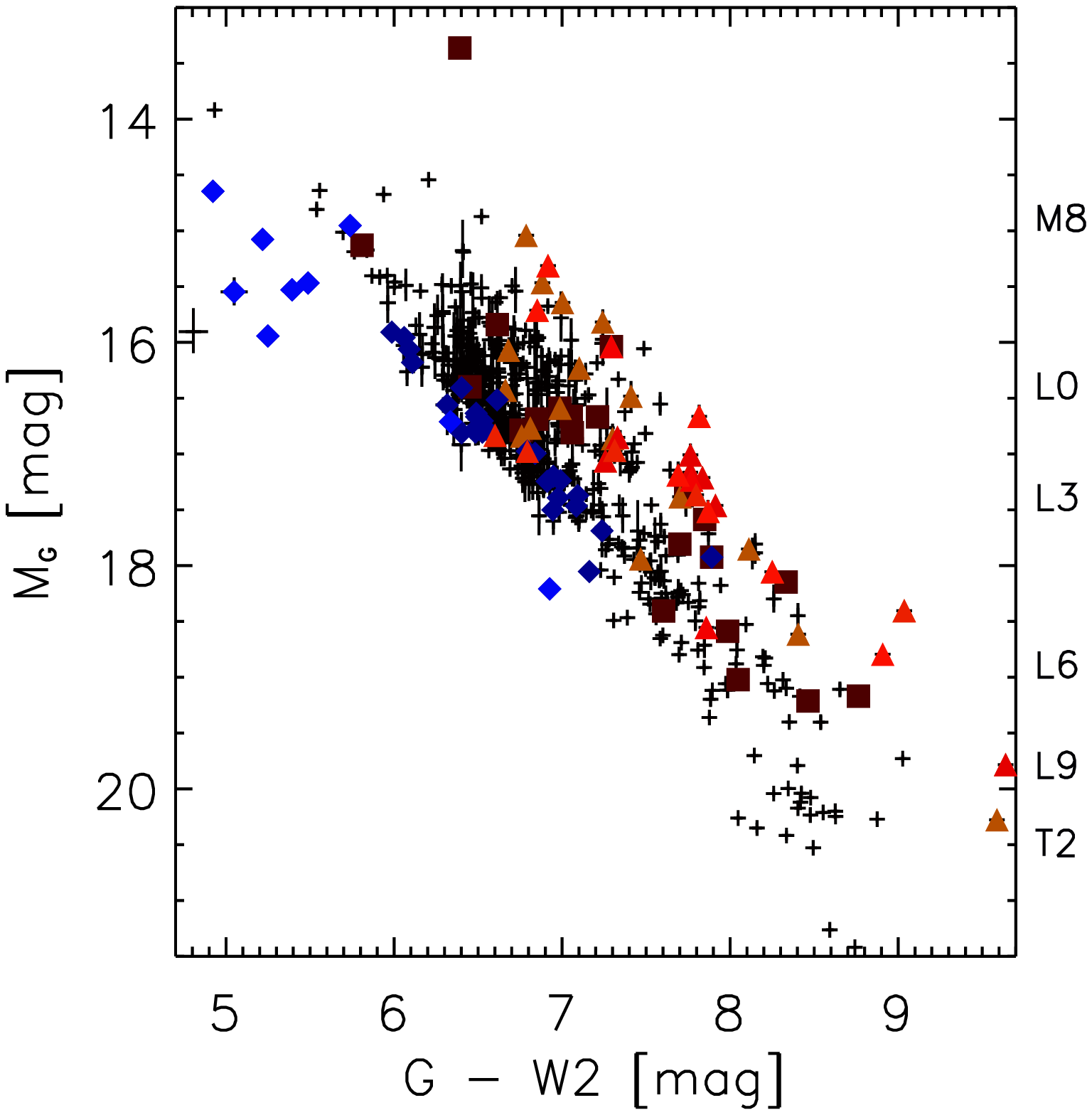}
\includegraphics[scale=0.43]{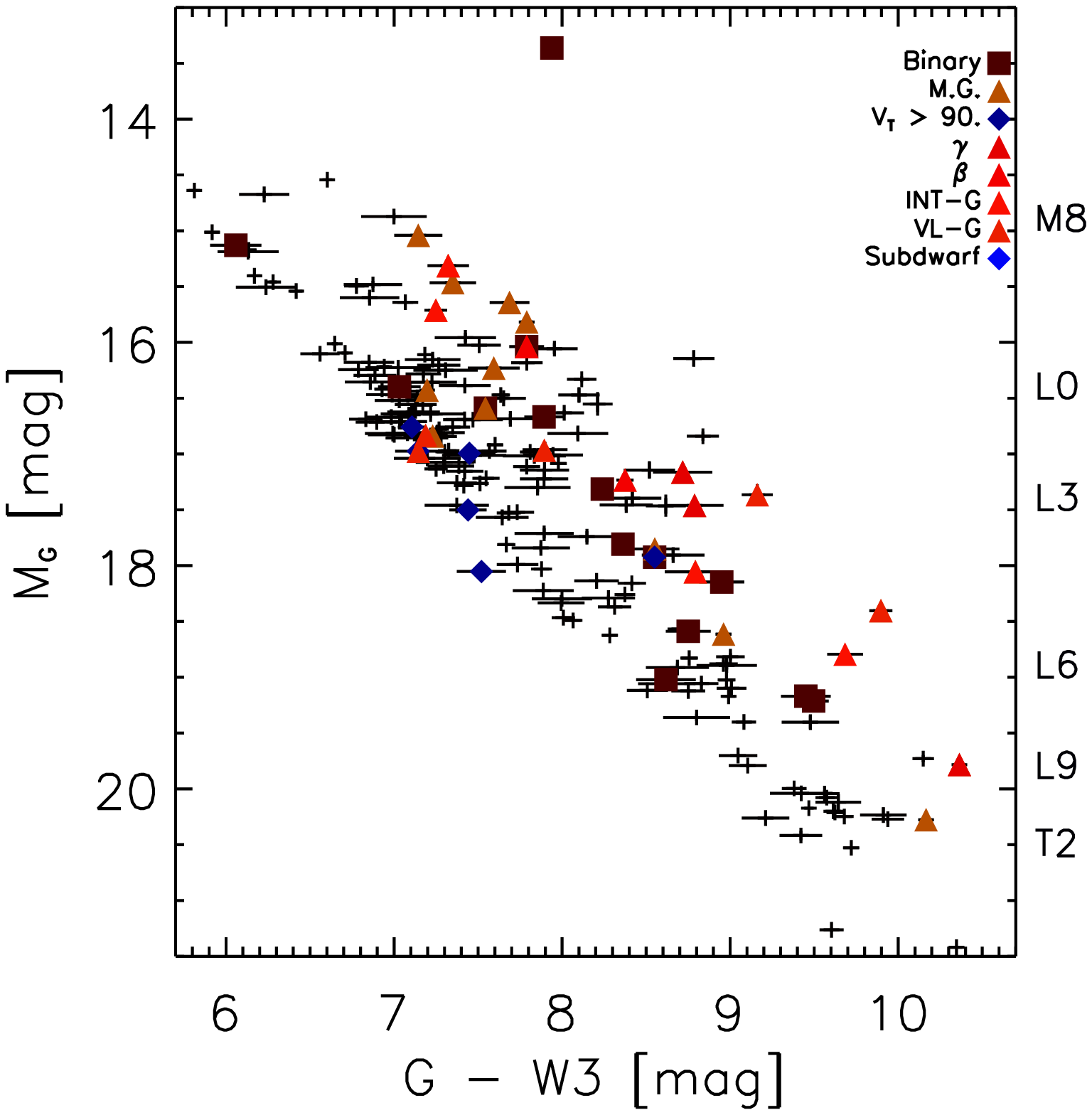}
\caption{Same as Figure~\ref{GUNNCMD} but for  $G-W1$, $G-W2$ and $G-W3$ colours from the
  {\em WISE} AllWISE catalogue. }
\label{WISECMD}
\end{center}
\end{figure}

The $G$--AllWISE colour-magnitude diagrams are marked by a drop in
objects with $WISE$ magnitudes (\NGOODWISEWA\ in $W1$, \NGOODWISEWB\ in $W2$,
\NGOODWISEWC\ in $W3$).  The sharper lower bound in
the main sequence of $G-W3$ indicates that the $J, H, K, W1, W2$ bands are
more complete than $\G$ for these objects, while the $W3$ band is incomplete.
The blueward trend for late L and T dwarfs seen in $G-K_s$ is still evident
in  $G-W1$, but in $G-W2$ and $G-W3$ the trend turns again redward, indicating
that temperature begins to dominate the spectral energy distribution as it
does in stars with spectral types M or earlier. 

The width of the main sequence in the $M_G$ vs. $G$--AllWISE colour-magnitude
diagrams continues to increase as it does in the $M_G$--2MASS ones.  The
mean $G$--AllWISE colours vary by $\sim$2.5\,mag and their dispersions
vary from 0.4 to 0.6\,mag. The mean colours of the old and young samples
separate by 1.4\,mag and have dispersions of 0.7\,mag. The shallower depth of
the $W3$ band makes it impossible to quantitatively characterise the $G-W3$
colour, but the visible trends are consistent with a continuation of an
increasing dispersion and separation.

\subsubsection{Colour-magnitude diagrams summary}

The remarkably tight sequence in many field objects for $M_G$=17.5 to 19\,mag
seen in the Figure~\ref{GUNNCMD}, $M_G$ vs. $G-z$, reappears in the
Figure~\ref{TMASSCMD}, $M_G$ vs. $G-H$ and $G-K_S$, manifesting as two
distinct sequences for the youngest and oldest objects. There is a notable
concentration of older objects that have largely cooled to follow a relatively
narrow temperature versus luminosity sequence. The existence of these
sequences and the range of objects between them, which presumably have a
younger age or are binaries, is best illustrated by the $M_G$ vs. $G-W2$ and
$G-W3$ plots in the Figure~\ref{WISECMD}, where the scatter of the subdwarfs
and T dwarfs is markedly reduced. The 1.5 to 2 magnitudes of spread in $G-W3$
colour for a given $M_G$ for the whole spectral range through late M dwarfs, L
dwarfs and T dwarfs would likely make this the most useful diagnostic, though
the increased errors and lack of depth of $W3$ magnitudes and consequent loss
of subdwarfs limits the utility of this colour.


\subsection{Spectral type-\DR\ magnitude diagrams}
\begin{figure}
\begin{center}
\includegraphics[scale=0.5]{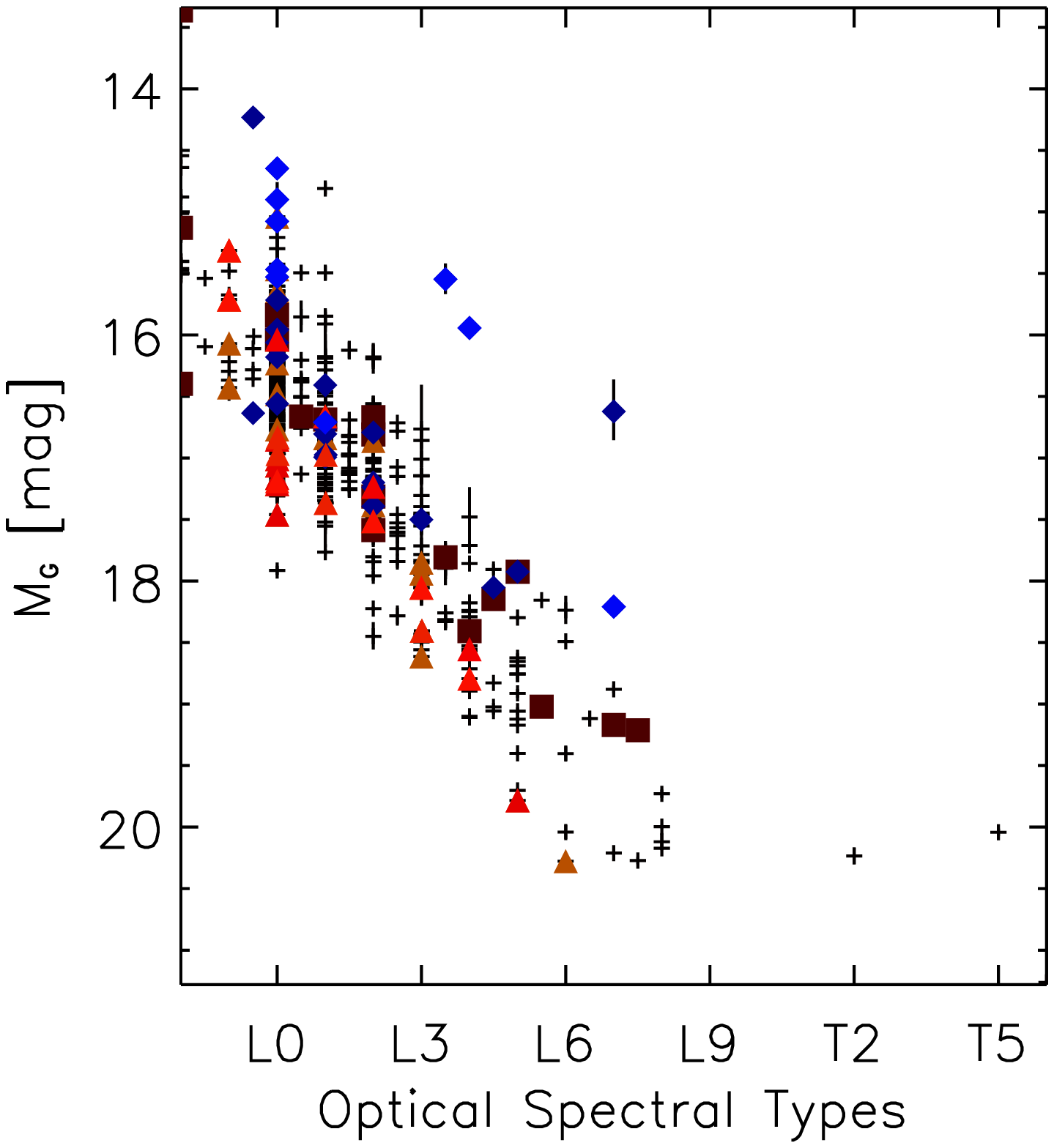}
\includegraphics[scale=0.5]{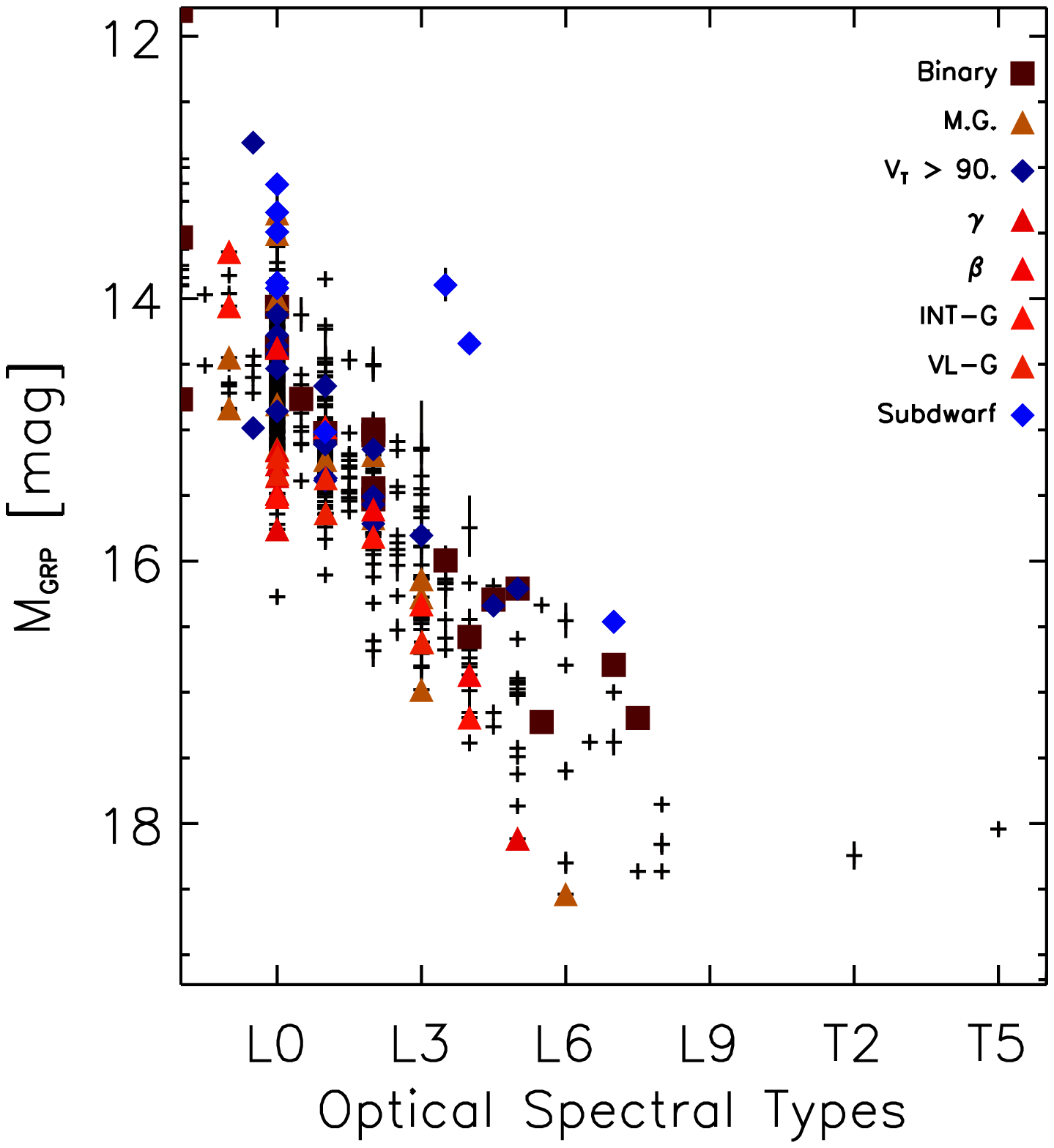}
\caption{Absolute magnitude in $G$ (top) and $\grp$ (bottom) passbands
  vs. optical spectral types. } 
\label{SPTPLOTS}
\end{center}
\end{figure}

In Figure~\ref{SPTPLOTS} we plot the absolute magnitudes in the $G$ and
$\grp$ bands vs. published optical spectral types. For un-resolved objects the
observed spectral type is that of the brightest component, so it reflects the temperature
of only that component. If it is an equal mass system the observed spectral
type is the approximate type of both components.
For these passbands the subdwarfs tend to appear overluminous while the
younger objects underluminous. The binaries are in general overluminous as
the spectral type is the temperature of only one component, while the
magnitude has a contribution from both components. The differences are not
always consistent because the contributions of the secondaries vary. However,
the $\grp$ magnitudes are more offset than the $G$ magnitudes due to those
estimates being the combined values instead of the component values. In light
of our discussion in Section~\ref{EPPROB} about problems with $\gbp$ and the
higher signal to noise of $\grp$, in future \G\ data releases it might well be
appropriate to make comparisons using $\grp$ rather than $G$. We have also
made similar plots comparing to the spectral types determined from the near
infrared colours (not shown) and the relations are similar to those shown,
though with larger spreads.


\subsection{Absolute Magnitude Relations}

There have been many determinations of the relation between absolute magnitude
and spectral types. For M, L and T dwarfs this has been derived as a simple
polynomial fit to a sample of classified objects with measured parallaxes and
apparent magnitudes \cite[e.g. for M, L and T dwarfs
  see][]{2002AJ....124.1170D, 2004AJ....127.2948V, 2004AJ....128.2460H,
  2012ApJS..201...19D, 2013AJ....146..161M}. While the number of objects per
spectral class bin was small and the relative error of the parallaxes was
large, such a simple approach was justified. The \G\ LT dwarf sample is,
especially for early L dwarfs, large and the relative error of the
\DR\ parallaxes are small so this approach is no longer sufficient.

The determination of an absolute magnitude calibration is not straight forward
and there are a number of pitfalls:
\begin{itemize}[leftmargin=0.3cm,noitemsep,topsep=0pt]
\item when using the parallax with assumed Gaussian uncertainties to determine
  the absolute magnitude the resulting uncertainties in magnitude are
  non-Gaussian \citep{1973PASP...85..573L, 2003MNRAS.338..891S,
    2015PASP..127..994B,2018A&A...616A...9L};
\item the use of a magnitude limited sample leads to Malmquist-like biases;
\item young and old objects within the same spectral class have
  absolute  magnitudes that are systematically different, biasing the results;
\item close unresolved binaries bias the calibration to brighter magnitudes;
\item there is no physical reason to assume that the absolute magnitude and
  spectral types are related by a smooth polynomial.
\end{itemize}
Some of these problems can be alleviated by assuming an absolute magnitude vs.
colour relation \citep[e.g. ][]{2011AJ....141...98B}, but the use of colour
introduces other problems such as the inflection in colour at the L-T
transition \citep{2003AJ....126..975T}.

\begin{figure}
\begin{center}
\includegraphics[scale=0.45]{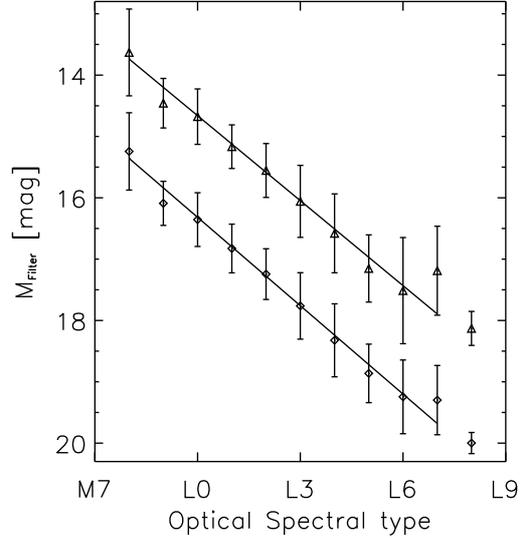}
\caption{Absolute $G$ (triangles) and $\grp$ (diamonds) magnitudes vs. optical
  spectral types. The points represent the medians as reported in
  Table~\ref{plotcalibrationtab}, the lines are the straight line fit to
  observations of M8 to L6 objects for the two passbands with parameters
  reported in Table~\ref{linear}.}
\label{plotcalibrationfig}
\end{center}
\end{figure}

For the \DR\ passbands we find the absolute magnitudes as a function of
optical spectral type for the bins where we have four or more objects, as
presented in Table~\ref{plotcalibrationtab}. The points in
Figure~\ref{plotcalibrationfig} are the inferred median absolute magnitudes
per spectral type calculated taking into account that the uncertainties in the
absolute magnitudes are neither Gaussian nor symmetric. The medians were
obtained using a Bayesian hierarchical model assuming that within each
spectral type bin there is a natural spread due to evolution and other
effects (for example metallicity), and an additional scatter due to the
observational uncertainties in the apparent magnitude and parallax. The
comparison with the observations that yield the likelihood term is done in the
space of parallaxes and apparent magnitudes. No distance estimation is involved
and no smoothness constraint is enforced in the model.

As can be seen in Figure \ref{plotcalibrationfig} the relation between M8 and
L6 is linear, which is true for the other passbands. The number of objects in
the other passbands is lower and the apparent magnitude precision is worse, so
separate absolute magnitude estimates for each spectral bin is not
warranted. Over this spectral range the error of a linear fit is smaller than
the scatter, so to enable absolute magnitude estimates as a function of
spectral type we made robust linear fits to all \Gcat\ objects with published
magnitudes of the form:
\begin{equation} M_{\lambda} = a_{\lambda} + b_{\lambda}~{\rm SpT}\end{equation}
valid in the range SpT=68 (M8) to 76 (L6). In Table~\ref{linear} we present
the parameters for the linear fits for all passbands. We include for
completeness the \DR\ passbands, though we recommend using calibration in
Table~\ref{plotcalibrationtab} for the most precise absolute magnitude
estimates.

\begin{table}
\caption{Absolute $G$ magnitude \DR\ calibration.}
\label{plotcalibrationtab}
\begin{center}
 \begin{tabular}{lrcc}
\\ 
\hline                                                
SpT   &  N   & $M_G$              & $M_{R_{p}}$  \\
Opt   &      &   [mag]                 &    [mag]       \\ 
\hline                                      
M8 &  16 & 15.24$\pm$0.63 & 13.63$\pm$0.71 \\ 
M9 &  17 & 16.09$\pm$0.36 & 14.46$\pm$0.40 \\ 
L0 & 234 & 16.36$\pm$0.44 & 14.68$\pm$0.45 \\ 
L1 & 103 & 16.83$\pm$0.40 & 15.17$\pm$0.36 \\ 
L2 &  68 & 17.24$\pm$0.41 & 15.55$\pm$0.44 \\ 
L3 &  41 & 17.76$\pm$0.54 & 16.06$\pm$0.59 \\ 
L4 &  26 & 18.32$\pm$0.60 & 16.58$\pm$0.64 \\ 
L5 &  17 & 18.86$\pm$0.48 & 17.15$\pm$0.55 \\ 
L6 &   6 & 19.25$\pm$0.60 & 17.51$\pm$0.87 \\ 
L7 &   8 & 19.30$\pm$0.56 & 17.19$\pm$0.73 \\ 
L8 &   4 & 20.00$\pm$0.17 & 18.13$\pm$0.28 \\ 

\hline
\end{tabular}
\end{center}
\end{table}

\begin{table}
\caption{Linear fits to absolute magnitude and spectral type for different
  passbands. }
\label{linear}
\begin{center}
\begin{tabular}{lccc}
\hline                                                
Absolute   &  $N$   & $a_{\lambda}$   & $b_{\lambda}$  \\ 
magnitude  &        &  [mag]       & [mag] \\ 
\hline                                      
 &  &  Optical SpT \\
 M$_{\rm G}$ & 477   & -17.303$\pm$  0.568   &   0.480$\pm$  0.004    \\%
     M$_{r}$ & 323   & -11.159$\pm$  1.977   &   0.419$\pm$  0.033    \\%
     M$_{i}$ & 380   & -18.351$\pm$  0.771   &   0.489$\pm$  0.008    \\%
 M$_{R_{p}}$ & 455   & -17.663$\pm$  1.162   &   0.462$\pm$  0.016    \\%
     M$_{z}$ & 356   & -18.001$\pm$  1.436   &   0.463$\pm$  0.013    \\%
     M$_{y}$ & 358   & -17.957$\pm$  0.627   &   0.449$\pm$  0.003    \\%
     M$_{J}$ & 475   & -14.479$\pm$  0.408   &   0.373$\pm$  0.005    \\%
     M$_{H}$ & 444   & -11.304$\pm$  0.750   &   0.317$\pm$  0.010    \\%
   M$_{K_s}$ & 442   &  -9.342$\pm$  0.568   &   0.282$\pm$  0.008    \\%
    M$_{W1}$ & 438   &  -4.983$\pm$  1.036   &   0.216$\pm$  0.008    \\%
    M$_{W2}$ & 435   &  -4.008$\pm$  0.955   &   0.198$\pm$  0.010    \\%
    M$_{W3}$ & 422   & -11.554$\pm$  0.878   &   0.292$\pm$  0.012    \\%
\hline
&  & Infrared SpT  \\
 M$_{\rm G}$ & 319   &  -7.717$\pm$  0.979   &   0.347$\pm$  0.017    \\%
     M$_{r}$ & 197   &   0.446$\pm$  2.305   &   0.257$\pm$  0.033    \\%
     M$_{i}$ & 254   & -10.941$\pm$  1.488   &   0.386$\pm$  0.016    \\%
 M$_{R_{p}}$ & 300   &  -8.294$\pm$  1.545   &   0.331$\pm$  0.023    \\%
     M$_{z}$ & 234   & -10.109$\pm$  1.398   &   0.353$\pm$  0.022    \\%
     M$_{y}$ & 239   &  -9.706$\pm$  1.776   &   0.334$\pm$  0.026    \\%
     M$_{J}$ & 317   &  -8.001$\pm$  0.602   &   0.282$\pm$  0.009    \\%
     M$_{H}$ & 313   &  -5.238$\pm$  1.460   &   0.232$\pm$  0.018    \\%
   M$_{K_s}$ & 314   &  -4.367$\pm$  1.259   &   0.213$\pm$  0.019    \\%
    M$_{W1}$ & 309   &  -0.232$\pm$  0.575   &   0.150$\pm$  0.008    \\%
    M$_{W2}$ & 308   &   0.051$\pm$  0.868   &   0.142$\pm$  0.014    \\%
    M$_{W3}$ & 299   &  -5.085$\pm$  0.948   &   0.202$\pm$  0.014    \\%
\hline

\end{tabular}
\end{center}

Parameters for Equation 3: $M_{\lambda} = a_{\lambda} + b_{\lambda}~{\rm
  SpT}$, valid in the range M8 to L6. The top set of parameters applies when
using optical spectral types and the lower set for infrared
spectral types. \end{table}

\section{Conclusion}
We have searched for known ultra-cool dwarfs in the \DR\ and found \NINCAT\ with
measured parallaxes, proper motions and $G$ magnitudes. We have matched this
dataset to publicly available large optical and infrared surveys, and
produced a catalogue that we make available to the community and will use as a
training set in the \G\ data processing chain of the ultra-cool dwarfs work
package. We have discovered \NBINNEWFOUND\ new multiple systems in our LT
catalogue. We have examined a number of colour-magnitude diagrams finding
significant main sequence structure in the ultra-cool dwarf region. We find
the $\gbp$ magnitude is not reliable for this sample and caution against using
it for selection and interpretation.

We are currently using this sample to develop and refine procedures for a
large scale search of the full \DR\ to discover previously unknown ultra-cool
dwarfs. We expect there to be over 300 new LT dwarfs and there will be
1000s of new late M-type ultra-cool dwarfs. We will catalogue and
examine in an automatic way these new objects looking for fine structure in
the spectro-photometric trends and find outlier objects that will
indicate new physical processes or environments.


\section*{Acknowledgments}

The authors thank the anonymous referee for a thorough review that increased
the quality of this contribution.
We thank Jonathan Gagn\'e for useful
discussions during the preparation of this manuscript.
DB was supported by Spanish grant ESP2015-65712-C5-1-R; JCB by Proyecto
FONDECYT postdoctorado 2018 nro. 3180716; FM by the NASA Postdoctoral Program
at the Jet Propulsion Laboratory, administered by Universities Space Research
Association under contract with NASA; JAC by Spanish grant
AYA2016-79425-C3-2-P; HRAJ by the UK’s Science and Technology Facilities
Council grant number ST/M001008/1.

This publication makes use of reduction and data products from the 
Centre de Donn\'ees astronomiques de Strasbourg (SIMBAD, \url{cdsweb.u-strasbg.fr}); 
the ESA \G\ mission (\url{gea.esac.esa.int/archive/}) funded by national institutions participating in the  \G~Multilateral
Agreement and in particular the support of ASI under contract I/058/10/0 (\G~
Mission - The Italian Participation to DPAC); 
the Panoramic Survey Telescope and Rapid Response System (Pan-STARRS, \url{panstarrs.stsci.edu}); 
the Sloan Digital Sky Survey (SDSS, \url{www.sdss.org}); 
the Two Micron All Sky Survey (2MASS, \url{www.ipac.caltech.edu/2mass}) 
and the Wide-field Infrared Survey Explorer ($WISE$, \url{wise.ssl.berkeley.edu}).

\bibliographystyle{mnras}
\bibliography{refs} 

\end{document}